\documentclass[lettersize,journal]{IEEEtran}
\usepackage{amsfonts}
\usepackage{amsmath}
\usepackage{algorithmic}
\usepackage{array}
\usepackage{colortbl}
\usepackage{multirow}
\usepackage[caption=false,font=normalsize,labelfont=sf,textfont=sf]{subfig}
\usepackage{textcomp}
\usepackage{stfloats}
\usepackage{tabularx}
\usepackage{booktabs}
\usepackage{url}
\usepackage{float}
\usepackage{tikz,xcolor}

\usepackage{times}
\usepackage{latexsym}
\usepackage{amssymb}

\newcommand{\mcL}{\mathcal{L}}

\usepackage[pagebackref,breaklinks,colorlinks]{hyperref}
\usepackage{hyperref}
\hypersetup{
    colorlinks=true,
    linkcolor=blue,
    filecolor=blue,      
    urlcolor=blue,
    citecolor=cyan,
}

\usepackage{amsmath,amsfonts,bm}

\def\eqref#1{equation~\ref{#1}}

\def\1{\bm{1}}

\def\vy{{\bm{y}}}

\DeclareMathAlphabet{\mathsfit}{\encodingdefault}{\sfdefault}{m}{sl}
\SetMathAlphabet{\mathsfit}{bold}{\encodingdefault}{\sfdefault}{bx}{n}

\definecolor{lime}{HTML}{A6CE39}
\DeclareRobustCommand{\orcidicon}{
\begin{tikzpicture}
\draw[lime, fill=lime] (0,0)
circle[radius=0.16]
node[white]{{\fontfamily{qag}\selectfont \tiny \.{I}D}};
\end{tikzpicture}
\hspace{-2mm}
}
\foreach \x in {A, ..., Z}{%
\expandafter\xdef\csname orcid\x\endcsname{\noexpand\href{https://orcid.org/\csname orcidauthor\x\endcsname}{\noexpand\orcidicon}}
}

\def\BibTeX{{\rm B\kern-.05em{\sc i\kern-.025em b}\kern-.08em
    T\kern-.1667em\lower.7ex\hbox{E}\kern-.125emX}}
\usepackage{balance}

\begin{document}
\title{Advancing Grounded Multimodal Named Entity Recognition via LLM-Based Reformulation and Box-Based Segmentation}

\author{Jinyuan Li, Ziyan Li, Han Li, Jianfei Yu, Rui Xia, Di Sun, and Gang Pan\IEEEauthorrefmark{1} 
\thanks{Jinyuan Li and Gang Pan are with the College of Intelligence and Computing, Tianjin University, Tianjin, China. Ziyan Li, Jianfei Yu, and Rui Xia are with NJUST, Nanjing, China. Han Li is with the College of Mathematics, Taiyuan University of Technology, Taiyuan, China. Di Sun is with Tianjin University of Science and Technology, Tianjin, China.}
\thanks{This work was supported by National Natural Science Foundation of China (62576243) and National Key Research and Development Program of China (2024YFB4710704).}
\thanks{\IEEEauthorrefmark{1}Corresponding author: Gang Pan. e-mail: pangang@tju.edu.cn}
}


\maketitle

\begin{abstract}
Grounded Multimodal Named Entity Recognition (GMNER) task aims to identify named entities, entity types and their corresponding visual regions. GMNER task exhibits two challenging attributes: 1) The tenuous correlation between images and text on social media contributes to a notable proportion of named entities being ungroundable. 2) There exists a distinction between coarse-grained noun phrases used in similar tasks (\emph{e.g.}, phrase localization) and fine-grained named entities. In this paper, we propose RiVEG, a unified framework that reformulates GMNE\underline{R} \underline{i}nto a joint modeling paradigm spanning MNER, \underline{VE}, and V\underline{G} perspectives by leveraging large language models (LLMs) as connecting bridges. This reformulation brings two benefits: 1) It enables us to optimize the MNER module for optimal MNER performance and eliminates the need to pre-extract region features using object detection methods, thus naturally addressing the two major limitations of existing GMNER methods. 2) The introduction of Entity Expansion Expression module and Visual Entailment (VE) module unifies Visual Grounding (VG) and Entity Grounding (EG). This endows the proposed framework with unlimited data and model scalability. Furthermore, to address the potential ambiguity stemming from the coarse-grained bounding box output in GMNER, we further construct the new Segmented Multimodal Named Entity Recognition (SMNER) task and corresponding Twitter-SMNER dataset aimed at generating fine-grained segmentation masks, and experimentally demonstrate the feasibility and effectiveness of using box prompt-based Segment Anything Model (SAM) to empower any GMNER model with the ability to accomplish the SMNER task. Extensive experiments demonstrate that RiVEG significantly outperforms SoTA methods on four datasets across the MNER, GMNER, and SMNER tasks. Datasets and Code will be released at \url{https://github.com/JinYuanLi0012/RiVEG}. 
\end{abstract}


\section{Introduction}
\IEEEPARstart{M}{ultimodal} Named Entity Recognition (MNER) on social media is a classical multimodal information extraction task~\cite{moon2018multimodal,zhang2018adaptive}. 
Due to the abundance of informal and brief unstructured textual content on social media platforms \cite{wang2019learning,chen2017predicting,ji2018cross}, many studies~\cite{lu2018visual,yu2020improving,zhang2021multi} attempt to improve Named Entity Recognition (NER) performance by leveraging multimodal features. Compared with traditional NER methods that solely consider text information~\cite{chiu2016named,huang2015bidirectional,ma2016end,kenton2019bert}, MNER approaches demonstrate superior performance in many scenarios~\cite{jia2022query,zheng2020object,wang2022cat,wang2022ita,wang2022named,zhao2022learning}. However, MNER only focuses on extracting entity-type pairs from image-text pairs and cannot fully address the pressing needs in related domains, such as building multimodal knowledge graphs~\cite{chen2023building} and enhancing human-computer interaction~\cite{xu2022skeleton}. 
Recent studies~\cite{yu2023grounded,wang2023fine} endeavor to conduct more in-depth extensions on MNER. As an emerging multimodal task, Grounded Multimodal Named Entity Recognition (GMNER) aims to extract text named entities and entity types from image-text pairs while performing visual grounding of named entities in form of bounding boxes. 

The existing GMNER method~\cite{yu2023grounded} uses object detection (OD) techniques to extract candidate region features from images. Then it builds a set of image and text multimodal fusion features through the end-to-end architecture and completes prediction of text named entities and entity-region matching based on these features. However, there are two obvious limitations to this method: 1) It compromises on MNER performance in order to endow the model with the capability to handle the visual task. 2) It heavily relies on OD models to pre-extract candidate region features, and these candidate regions may not always contain the ground truth visual region. This results in a natural performance ceiling for this method.

\begin{figure}[t] 
	\begin{center}
		\includegraphics[width=1\linewidth]{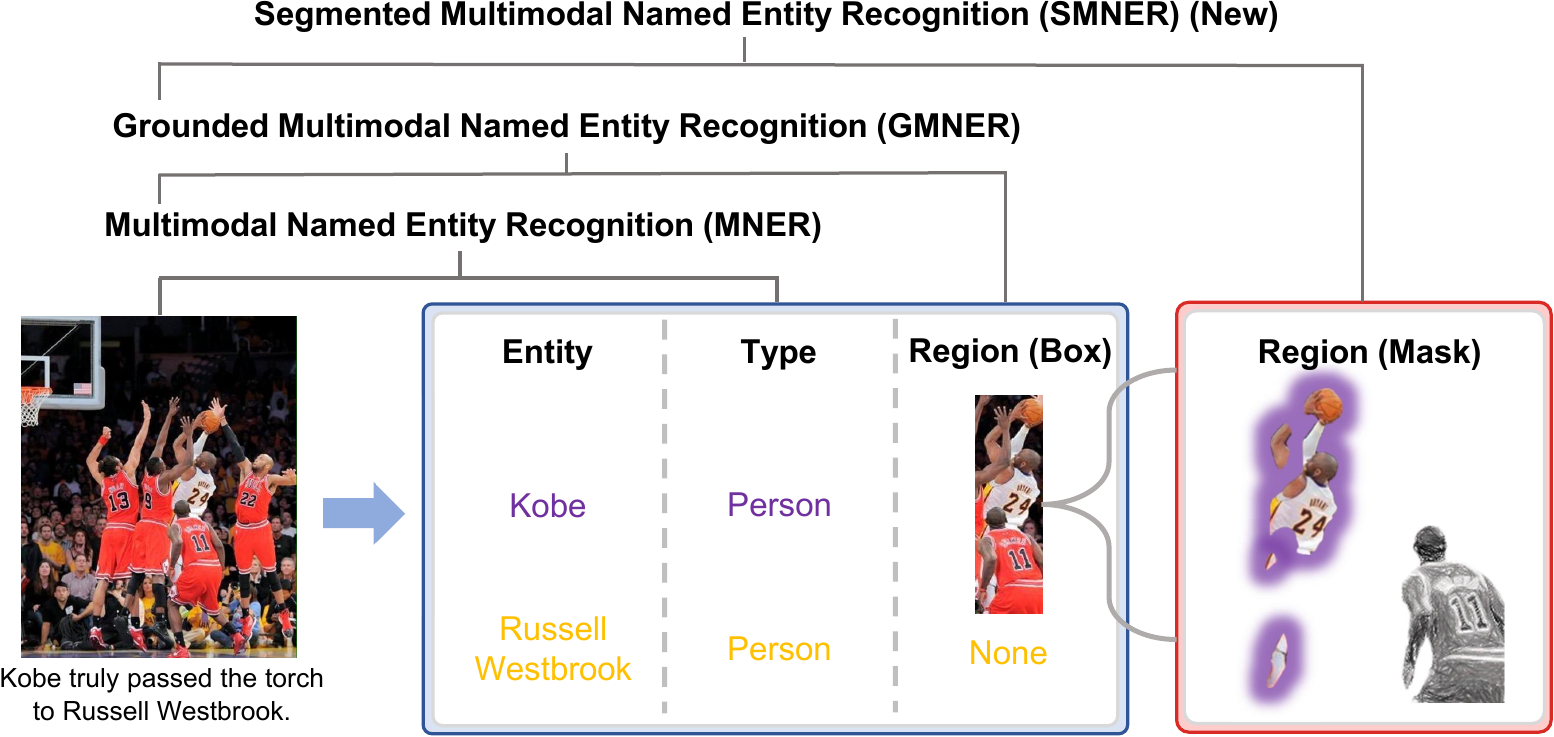}
	\end{center}
	\vspace{-4mm}
	\caption{Comparison of MNER, GMNER, and SMNER tasks. In the GMNER task, a single box may include multiple visual objects, obscuring fine-grained boundaries. The SMNER task addresses this issue by enabling finer-grained visual entity identification.}
	\label{fig:seg_intro} \vspace{-6mm}
\end{figure}

Moreover, given the intricate nature of images on social media, relying solely on coarse-grained bounding boxes to locate visual objects may not always yield effective results. As shown in Fig.~\ref{fig:seg_intro}, even when the existing GMNER model correctly grounds named entities using bounding boxes, these boxes may still include multiple similar visual objects. This ambiguity is detrimental to practical applications.

To address the aforementioned issues, we propose a unified framework that aims to leverage large language models (LLMs) as bridges to reformulate GMNER as a two-stage joint of MNER and Entity Grounding (EG). This reformulation directly solves two limitations of existing GMNER methods. The first stage aims to introduce external knowledge appropriately to ensure the optimal MNER performance. The second stage aims to introduce the Visual Grounding (VG) method to naturally bypass the limitations of the OD method and avoid using difficult entity-region matching. Furthermore, to address potential ambiguities in the grounding process, we further propose a new Segmented Multimodal Named Entity Recognition (SMNER) task and the corresponding dataset. This task aims to extract named entities and entity types while further predicting segmentation masks of visual objects. We highlight the main contributions as follows.


\textbf{Unleash the potential of LLMs in MNER by guiding LLMs to generate auxiliary refined knowledge of the original samples.} Rapidly developing LLMs achieve promising results in various NLP tasks~\cite{shao2023prompting,yang2022empirical}. But recent research~\cite{qin2023chatgpt,wang2023gpt} shows that LLMs exhibit shortcomings in sequence labeling tasks such as full-shot NER. Considering that the amount of data and task characteristics make it difficult to directly improve LLMs in MNER, we regard implicit knowledge bases such as LLMs as auxiliary tools and explore how to \underline{\textbf{P}}rompt \underline{\textbf{L}}LMs \underline{\textbf{I}}n \underline{\textbf{M}}NER (PLIM). Specifically, we construct a set of manually annotated samples and design a Multimodal Similar Example Awareness (MSEA) module. For any new input sample, similar manually annotated samples are picked out by the MSEA module for LLMs to perform in-context learning and generate auxiliary refined knowledge that helps understand the sample. Experimental results show that the proposed MNER module (PLIM) exploits the potential of LLMs and effectively addresses the limitations of traditional MNER methods~\cite{yu2020improving,wang2022ita,wang2022named}, surpassing current SoTA methods on two classic MNER datasets. 

\begin{figure}[t] 
	\begin{center}
		\includegraphics[width=1\linewidth]{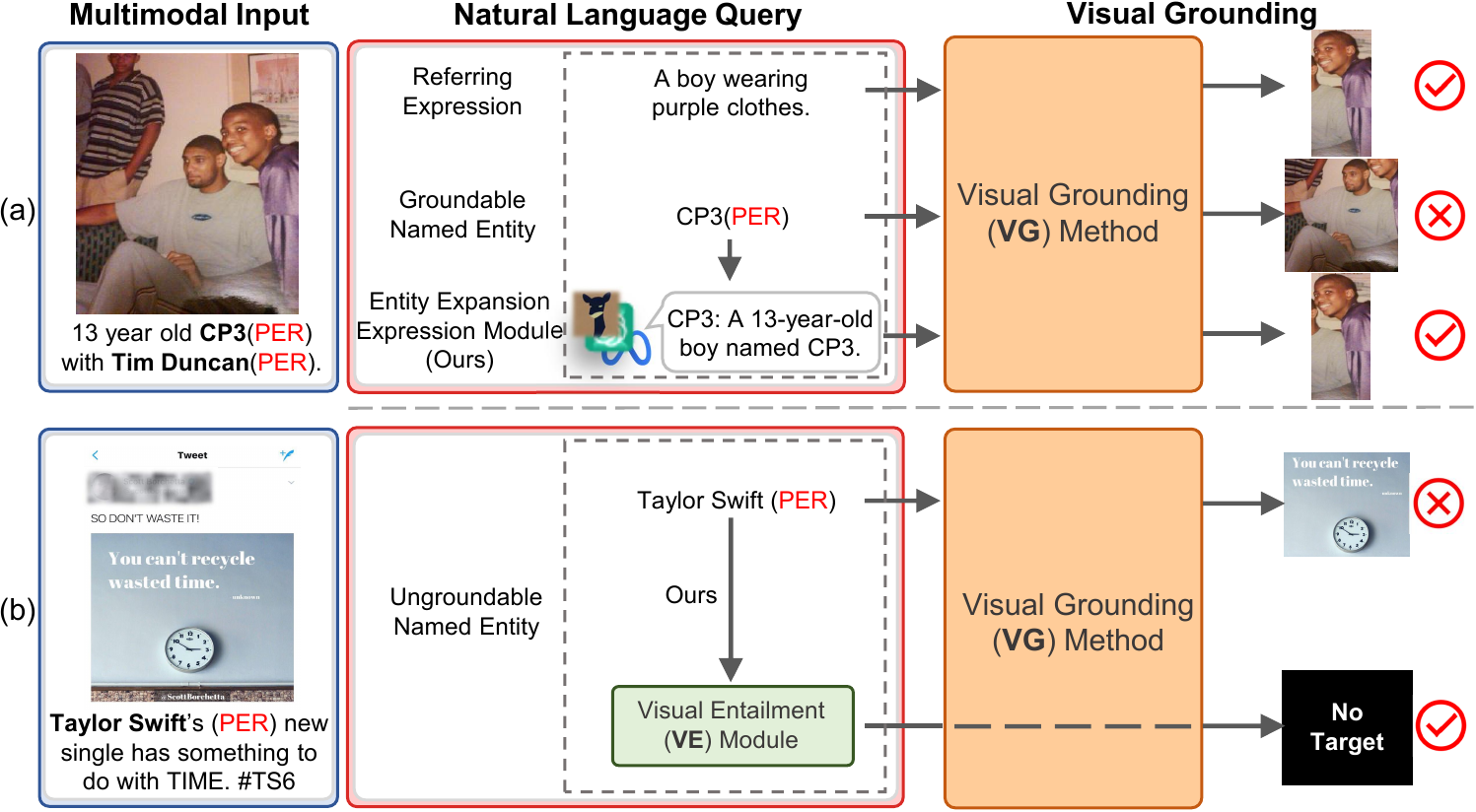}
	\end{center}
	\vspace{-4mm}
	\caption{\textcolor{black}{Traditional Visual Grounding methods are not suitable for real-world named entities or ungroundable text queries as input. To unify the Entity Grounding (EG) task and Visual Grounding (VG) methods, we introduce the Entity Expansion Expression module and the Visual Entailment (VE) module.}}
	\label{fig:VGandEG}
	\vspace{-4mm}
\end{figure}

\textbf{Unified Visual Grounding (VG) and Entity Grounding (EG) by introducing the LLM-based entity expansion expression and the Visual Entailment (VE) module.} The Entity Grounding stage in GMNER is defined as determining the groundability and grounding results of input named entities. And the traditional Visual Grounding task~\cite{rohrbach2016grounding,yu2016modeling} aims to perceive relevant regions in images based on input text queries. Many VG methods do not depend on OD techniques~\cite{zhu2022seqtr,liu2023polyformer,yan2023universal,wang2022ofa}, which means that unifying VG and EG can naturally overcome the limitation of the existing GMNER method~\cite{yu2023grounded} constrained by OD techniques. But there are two significant differences between VG and EG: 1) The text input of the VG task consists of noun phrases or referring expressions, which is essentially different from the real-world named entities of the EG task. In Fig.~\ref{fig:VGandEG}\textcolor{blue}{a}, the VG method can understand ``\textit{a boy wearing purple clothes}", but struggles with named entities like ``\textit{CP3}". 2) In Fig.~\ref{fig:VGandEG}\textcolor{blue}{b}, the traditional VG task only involves groundable text inputs and does not involve any ungroundable text inputs such as ``\textit{Taylor Swift}". This means that VG models always treat text queries as positive queries and attempt to output grounding results. However, due to the habits of users in utilizing social media, approximately 60\% of named entities in the GMNER dataset are ungroundable~\cite{yu2023grounded}. This means that the behavior of existing VG methods~\cite{zhu2022seqtr,liu2023polyformer,yan2023universal,wang2022ofa} is undefined for this scenario. In order to achieve the unification of VG and EG, we establish a bridge between the two based on the above differences. Specifically, we bridge the semantic gap between noun phrases and named entities by guiding LLMs to generate coarse-grained entity expansion expressions for fine-grained named entities, and introduce the Visual Entailment module to enable VG methods to cope with weakly correlated image-text pairs. Experimental results show that this unification enables the EG task to significantly benefit from existing VG research~\cite{wang2022ofa,li2021align} and helps train new SoTA GMNER models. 

\textbf{The entire framework is endowed with unlimited data scalability and model scalability by reformulating the task.} On the one hand, the existing 7k training data is limited for the realistic and challenging task like GMNER. 
However, once the GMNER task is reframed as a joint composition of modules grounded in MNER, VE, and VG, 
the entire framework can naturally inherit the corresponding pre-training foundations of different modules, \textit{e.g.,} about 6.1M language expressions and 174k images~\cite{yu2016modeling,nagaraja2016modeling,kazemzadeh2014referitgame,plummer2015flickr30k} available for the VG module pre-training~\cite{zhu2022seqtr}, about 14.1M image-text pairs~\cite{sharma2018conceptual,ordonez2011im2text,lin2014microsoft,krishna2017visual,changpinyo2021conceptual} available for the VE module pre-training~\cite{li2021align}, \textit{etc}. Furthermore, since different LLMs provide diverse auxiliary knowledge and entity expansion expressions for the same sample, the entire framework can further use various LLMs to enhance the limited 7k training data to gain data marginal benefits. 
On the other hand, wider training data supports better model scalability. The entire framework can naturally adapt to the relevant research and development of any single module to gain more model architecture advantages, and can freely select more accurate or lightweight independent modules based on application scenarios to constitute various variants. Experimental results show that all 14 variants of RiVEG exhibit better performance compared with existing baseline methods.

\textbf{The new task and dataset are constructed to achieve more fine-grained multimodal information extraction with almost no performance loss.} Since generating fine-grained segmentation masks of visual objects can effectively solve the ambiguity that may result from output bounding boxes, we propose a new Segmented Multimodal Named Entity Recognition (SMNER) task and construct a dataset that can be used for SMNER task by manually annotating the visual object segmentation masks of all groundable named entities in existing GMNER dataset. The SMNER task is more challenging than the GMNER task, because the former not only requires predicting the groundability of each entity but also predicting the fine-grained segmentation mask, which is more difficult than predicting the coarse-grained bounding box. Although it is challenging to directly build a SMNER model with acceptable performance based on about 4k groundable named entity samples in this dataset, the emergence of Segment Anything Model (SAM)~\cite{kirillov2023segment} makes it possible to solve this task. SAM cannot identify masked objects based on arbitrary text input. But considering that benefiting from prior knowledge of box positions and predicting the corresponding object masks based on bounding boxes can be more effective than directly predicting region masks based on text~\cite{zhang2023simple}, we integrate the capabilities of GMNER methods and SAM. With the entity groundability judgment and visual localization capabilities already possessed by GMNER methods, as well as the outstanding zero-shot segmentation ability of SAM based on box prompts, we demonstrate that any GMNER method can be equipped with the capability to accomplish SMNER task with almost no performance penalty.

This paper is a substantial extension of our previous Findings of EMNLP 2023~\cite{li2023prompting} and ACL 2024~\cite{li2024llms} works. This journal version has three major improvements: 

$\bullet$ We propose a new Segmented Multimodal Named Entity Recognition (SMNER) task, construct the corresponding Twitter-SMNER dataset, and provide detailed information about its construction process.

$\bullet$ To complete and evaluate the SMNER task, we effectively extend three existing GMNER methods~\cite{yu2023grounded} and the proposed RiVEG framework~\cite{li2024llms}. By leveraging the box prompt-based SAM~\cite{kirillov2023segment}, we demonstrate the feasibility of enabling any existing GMNER method to complete the SMNER task. This also allows us to discuss the adaptability of two different segmentation mask acquisition methods (\textit{i.e.}, text query-based and box prompt-based) for the SMNER task.

$\bullet$ We present more detailed and comprehensive quantitative and qualitative experiments on the PLIM module~\cite{li2023prompting} and the RiVEG framework~\cite{li2024llms}, including evaluations with various LLMs and text encoders, IoU threshold sensitivity analyses for the GMNER and SMNER tasks, as well as clearer and more intuitive case visualizations.

\section{Background}

\subsection{Terminology and Abbreviations}

Given the complexity of the proposed framework and the variety of tasks and modules involved, we summarize the main abbreviations in Table~\ref{tab:abbreviations} for clarity. These terms will appear frequently in the remainder of the paper.

\begin{table}[t]
\renewcommand{\arraystretch}{1.2}
\centering
\caption{List of abbreviations used in this work.}
\begin{tabular}{ll}
\hline
\textbf{Abbreviation} & \textbf{Full Term} \\
\hline
NER & Named Entity Recognition \\
MNER & Multimodal Named Entity Recognition \\
GMNER & Grounded Multimodal Named Entity Recognition \\
SMNER & Segmented Multimodal Named Entity Recognition \\
LLM & Large Language Model \\
OD & Object Detection \\
VG & Visual Grounding \\
VE & Visual Entailment \\
SAM & Segment Anything Model \\
PLIM & \underline{P}rompt \underline{L}arge Language Models \underline{I}n \underline{M}NER \\
MSEA & Multimodal Similar Example Awareness \\
RiVEG & Reformulating GMNE\underline{R} \underline{i}nto MNER, \underline{VE}, and V\underline{G} \\
\hline
\end{tabular}
\label{tab:abbreviations}
\end{table}

\subsection{Taxonomy of MNER Methods}
\begin{figure}[t] 
	\begin{center}
		\includegraphics[width=1\linewidth]{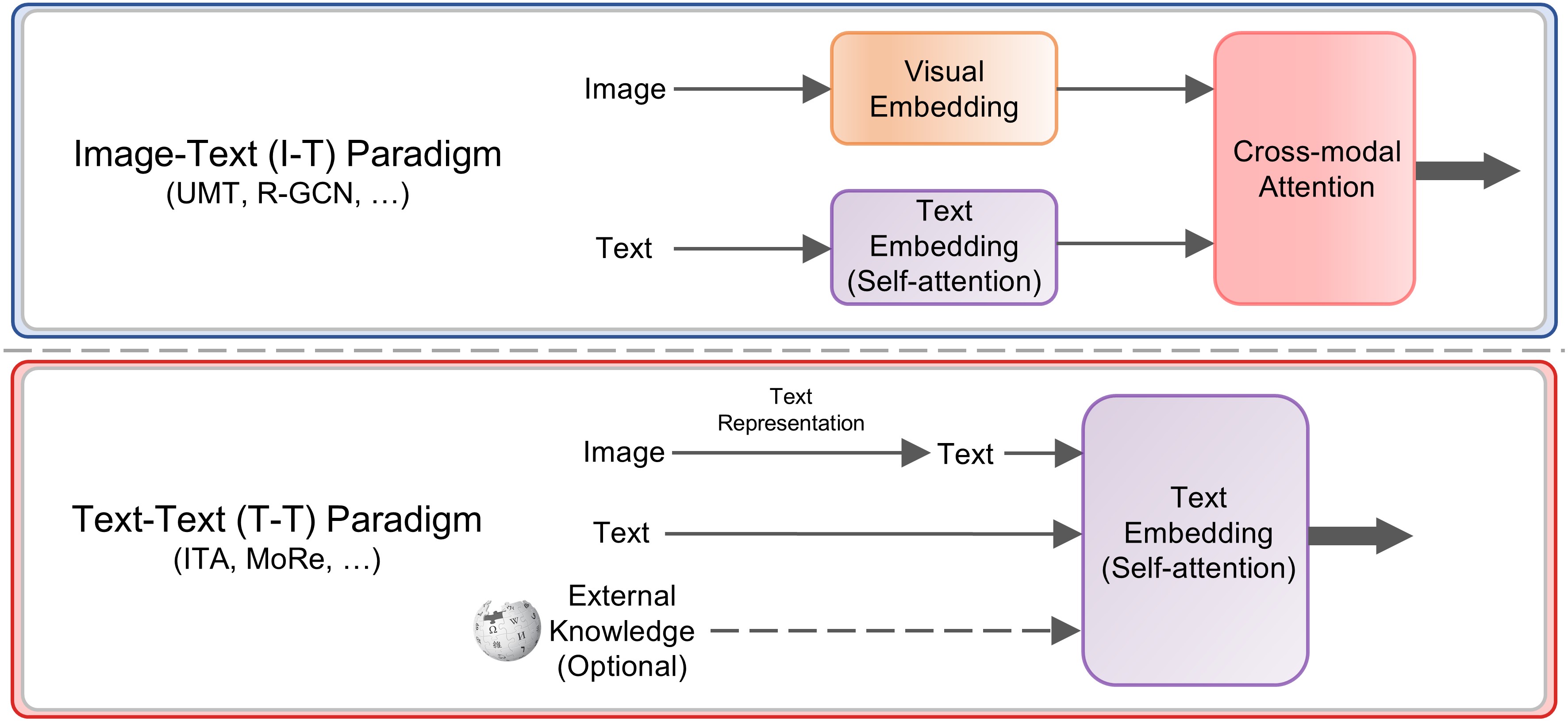}
	\end{center}
	\vspace{-4mm}
	\caption{\textcolor{black}{Two paradigms of existing MNER methods. The Text-Text (T-T) paradigm generally achieves better results than the Image-Text (I-T) paradigm because it avoids the cross-modal attention mechanism that is difficult to train.}}
	\label{fig:background}
	\vspace{-4mm}
\end{figure}

As shown in Fig.~\ref{fig:background}, existing MNER methods mainly manifest as the Image-Text (I-T) paradigm and the Text-Text (T-T) paradigm. Early methods mainly select the I-T paradigm~\cite{yu2020improving,zheng2020object,chen2023learning,zhao2022learning,wang2022cat}, aiming to combine text features with image information by designing various cross-modal attention mechanisms. However, these methods consistently constrained by following two limitations: 1) The image feature extractors (\textit{e.g.,} ResNet~\cite{he2016deep}, Mask R-CNN~\cite{he2017mask}) used by these methods are mainly trained based on datasets such as ImageNet~\cite{deng2009imagenet} and COCO~\cite{lin2014microsoft}. There is a significant semantic gap between the noun labels of these datasets and the named entities in real scenarios. 2) There are significant differences in the feature distributions of different modalities. And since the two classic MNER training sets only contain 4000~\cite{zhang2018adaptive} and 3373~\cite{lu2018visual} image-text pairs, this disparity is difficult to reconcile by training on a small amount of MNER data. 
Given these limitations, the performance of some I-T paradigm MNER methods is inferior to SoTA language models that only focus on text~\cite{wang2022ita}.

One particularity of the MNER task is that the introduction of image information is designed to assist in understanding text and eliminating ambiguities. This means that using implicit features to represent images is not the only way. Using text to describe images can still achieve the same auxiliary effect in most cases. And when images cannot provide more supplementary information for understanding the text, the introduction of external knowledge becomes a better supplement. Therefore, recent T-T paradigm research aims to solve MNER solely through text. ITA~\cite{wang2022ita} represents images using text extracted from images. MoRe~\cite{wang2022named} retrieves external textual knowledge from Wikipedia to assist the model in understanding text. Obviously, the self-attention mechanism within text modality is easier to train and more accurate than the cross-modal attention mechanism. Although this series of methods achieve more promising results than I-T paradigm methods, they still have shortcomings. Specifically, these methods either rely only on in-sample information and ignore external knowledge~\cite{wang2022ita}, or are too redundant to retrieve knowledge from external explicit knowledge bases~\cite{wang2022named}. Considering the significance of external knowledge and the potential noise introduced by low-correlation external knowledge, it is imperative to devise an appropriately balanced approach for acquiring external knowledge.

\subsection{Existing GMNER Methods}
The GMNER task is disassembled into the MNER stage and the Entity Grounding (EG) stage for analysis. The MNER stage predicts text named entities and corresponding entity types, while the EG stage receives the predictions from the MNER stage and determines the groundability and grounding results of named entities. The existing GMNER method~\cite{yu2023grounded} uses an end-to-end architecture to construct multimodal fusion features of images and text and completes MNER and EG based on this fusion features. But this method has two obvious limitations: 

$\bullet$ The MNER performance of this method is suboptimal. This aggravates the error propagation in the overall prediction process, as suboptimal MNER performance is more likely to result in errors in predicting triples. The existing method performs MNER and EG based on the same set of multimodal fusion features, which results in this feature representation being neither the most superior to the MNER nor the most superior to the EG. Therefore, it is necessary to split the prediction of named entities and the grounding of named entities into independent stages, and further improve the shortcomings of existing T-T paradigm MNER methods~\cite{wang2022ita,wang2022named} in MNER stage to ensure the best MNER performance. 

\begin{table}[t]
\footnotesize
\centering
\caption{TopN-Prec@0.5 scores of two widely-adopted Object Detection (OD) methods on Twitter-GMNER dataset. The existing GMNER method is limited by OD techniques. Because these visual candidate regions may not contain visual gold labels for named entities.}
\begin{tabular}{l|ccc}
\toprule
\textbf{OD Methods} & \textbf{Top-10} & \textbf{Top-15} & \textbf{Top-20}  \\
\midrule
Faster R-CNN~\cite{anderson2018bottom} & 59.87\% & 69.84\% & 76.11\%  \\
VinVL~\cite{zhang2021vinvl} & 66.69\%  & 74.62\% & 84.29\%   \\
\bottomrule
\end{tabular}
\label{tab:OD test} \vspace{-3mm}
\end{table}

$\bullet$ The existing GMNER method uses object detection (OD) methods~\cite{anderson2018bottom,zhang2021vinvl} to detect all visual objects in images, and then performs entity-region matching between named entities and all candidate visual objects. But as shown in Table~\ref{tab:OD test}, these classic OD methods perform poorly due to the complexity and diversity of images originating from social media.\footnote{Here, TopN-Prec@0.5 is the proportion of entities where at least one of the Top-N predicted bounding boxes based on detection probability has an IoU of 0.5 or greater with the ground truth bounding box.} This means that there is a natural performance ceiling for this kind of matching. On the one hand, samples that fail to detect ground truth visual regions by OD methods in the prediction stage inevitably have erroneous entity-region matching. On the other hand, the existing method treat named entity samples where ground truth visual regions are not successfully detected by OD methods as ungroundable samples in the training stage (\textit{i.e.}, 7088/4694 to 8262/3520 ungroundable/groundable). This exacerbates the data imbalance of the GMNER dataset~\cite{yu2023grounded}, potentially leading the model to bias towards predicting entities as ungroundable. Therefore, it is necessary to avoid such entity-region matching based on OD methods.

\begin{table}[t]
\footnotesize
\centering
\caption{Statistics of our proposed Twitter-SMNER dataset.}
\resizebox{0.42\textwidth}{!}{%
\begin{tabular}{l|cccc}
\toprule
\textbf{Split} & \textbf{\#Tweet}& \textbf{\#Entity} & \textbf{\#Groundable Entity}  & \textbf{\#Mask} \\
\midrule
Train & 7000 & 11782 & 4671 & 5581 \\
Dev   & 1500 & 2453  & 981  & 1163 \\
Test  & 1500 & 2543  & 1029 & 1229  \\
\midrule
Total & 10000  & 16778 & 6681 & 7973   \\
\bottomrule
\end{tabular}}
\label{tab:SMNER_dataset} \vspace{-4mm}
\end{table}

\section{Segmented Multimodal Named Entity Recognition Dataset} \label{sec:SMNER_dataset}

\subsection{Dataset Construction}
Since no dataset exists for the SMNER task, we construct a Twitter-SMNER dataset as follows. The existing GMNER dataset~\cite{yu2023grounded} accomplishes data collection, cleaning, and coarse-grained bounding box annotation of named entities. We further annotate fine-grained segmentation masks based on this foundation. 
Specifically, we employ five annotators and two experienced experts with research backgrounds in computer vision. The annotation tool and guidelines are also standardized to ensure consistency.\footnote{\url{https://github.com/labelmeai/labelme}} Each image-text pair is assigned to two different annotators. For controversial annotation samples, decisions are made by experienced experts. A tiny fraction of ambiguous samples is removed during the annotation process. Two annotations with IoU greater than 0.5 are considered consistent annotations. The Fleiss score between two different annotators is 0.82~\cite{fleiss1971measuring}. To further assess the consistency of the annotations, we also calculate the Dice Coefficient~\cite{dice1945measures}, which yields a score of 0.85. These results collectively demonstrate the high annotation consistency of the Twitter-SMNER dataset.

\begin{figure}[t] 
	\begin{center}
		\includegraphics[width=0.95\linewidth]{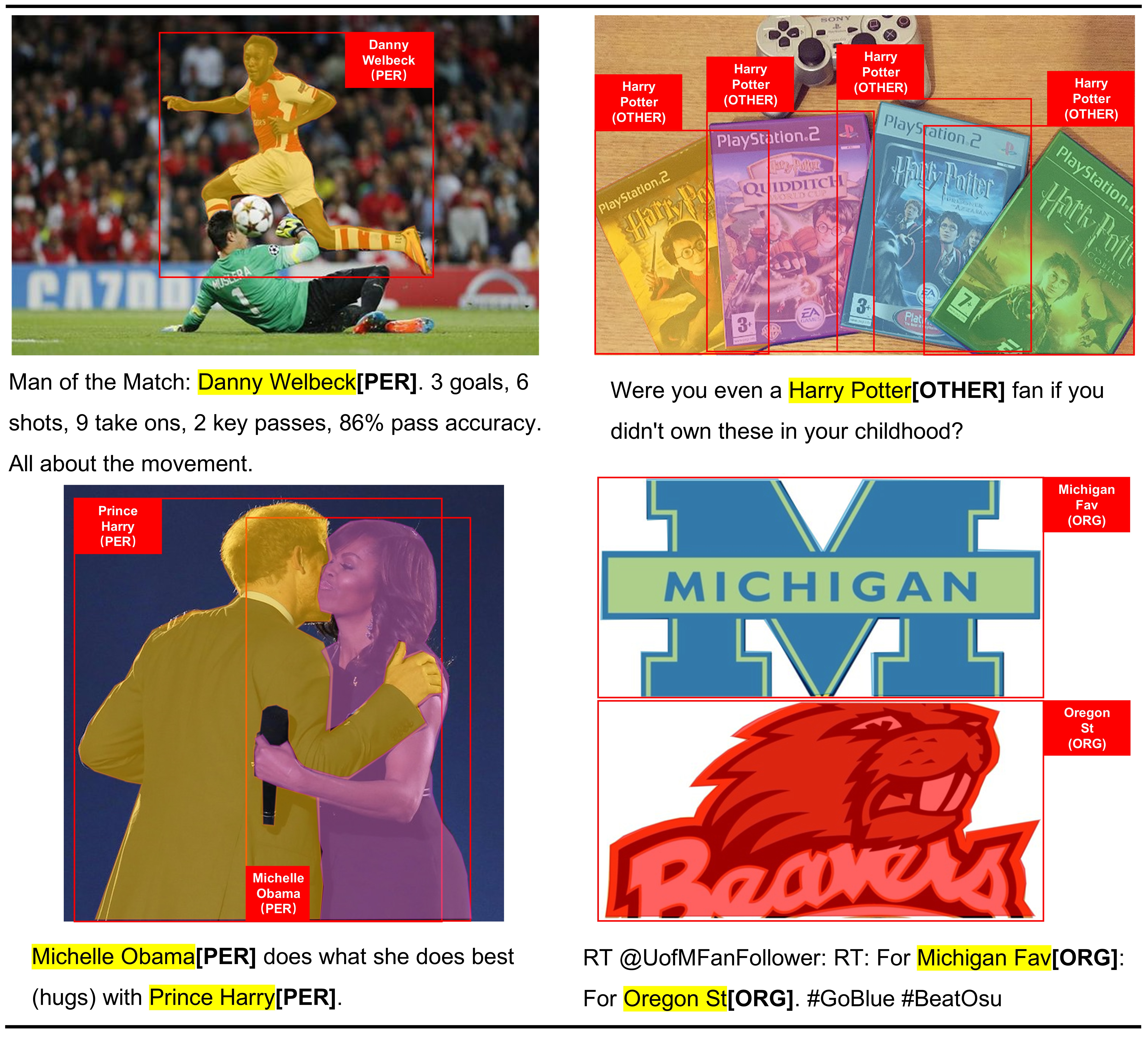}
	\end{center}
	\vspace{-4mm}
	\caption{Example annotations from the constructed Twitter-SMNER dataset.} 
	\label{fig:label example}
\end{figure}

In the data annotation process, we follow the standard visible instance segmentation setting that is widely adopted in datasets such as COCO~\cite{lin2014microsoft} and LVIS~\cite{gupta2019lvis}. As shown in Fig.~\ref{fig:label example}, each groundable entity is annotated with a separate segmentation mask that covers only the visible (non-occluded) regions. When occlusion occurs between different entities, their segmentation masks are allowed to overlap.

\subsection{Dataset Analysis}
\begin{figure}[t] 
	\begin{center}
		\includegraphics[width=1\linewidth]{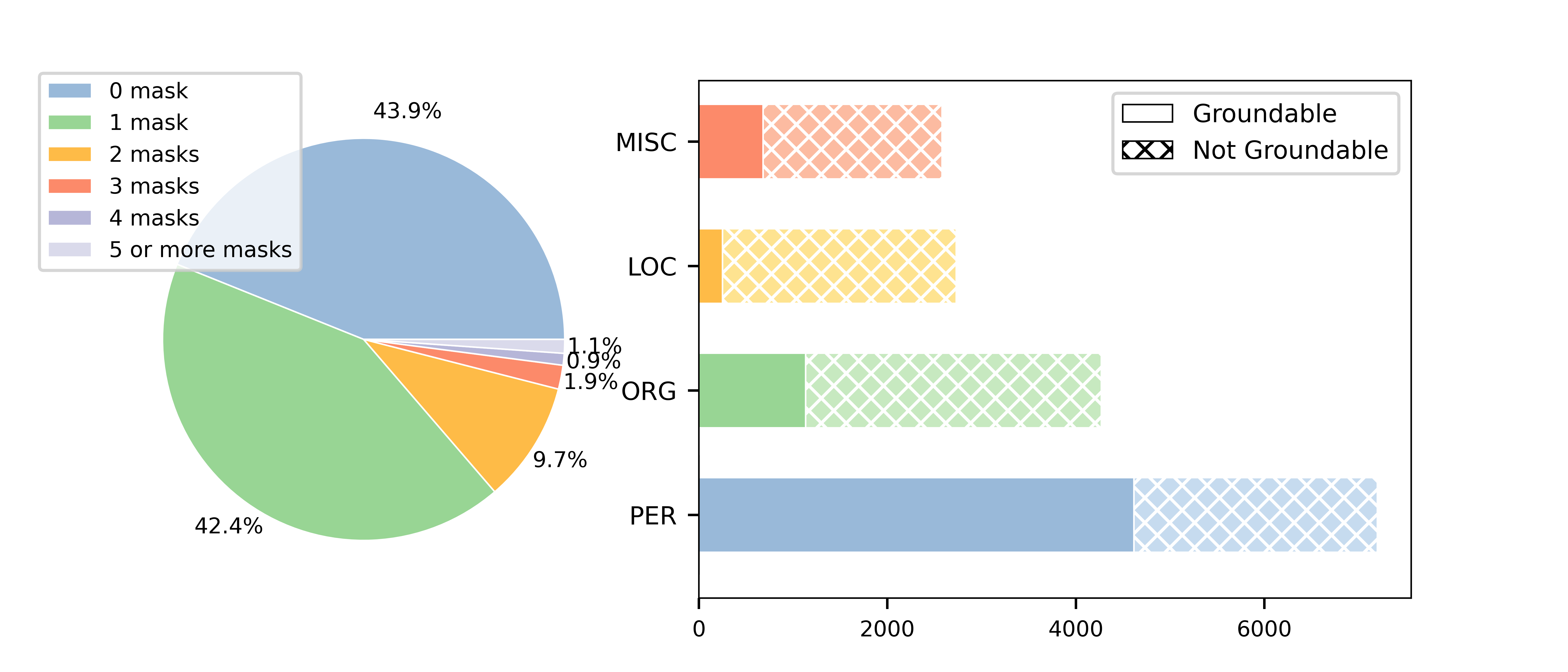}
	\end{center}
	\vspace{-4mm}
	\caption{\textcolor{black}{Distribution of the number of masks in each image and distribution of groundable and ungroundable entities for each entity type.}} \vspace{-2mm}
	\label{fig:dataset}
\end{figure}

We maintain the same partitioning as the Twitter-GMNER dataset~\cite{yu2023grounded}, \emph{i.e.}, 70\% is used for training set, 15\% for dev set, and 15\% for test set. As shown in Table~\ref{tab:SMNER_dataset}, the Twitter-SMNER dataset contains 16778 text named entities. For the 6681 groundable named entities, we annotate 7973 segmentation masks. This means that one entity may correspond to multiple segmentation masks. Fig.~\ref{fig:dataset} (left) shows that 43.9\% of the images do not contain any segmentation mask, 42.4\% of the images contain a single segmentation mask, and about 13.7\% of the images contain multiple segmentation masks. And Fig.~\ref{fig:dataset} (right) shows that named entities of the \textit{PER} type are often present in images, while the other three types are not. This reflects the challenge of the proposed SMNER task. 

\section{Methodology}

\begin{figure*}[t] 
	\begin{center}
		\includegraphics[width=1\linewidth]{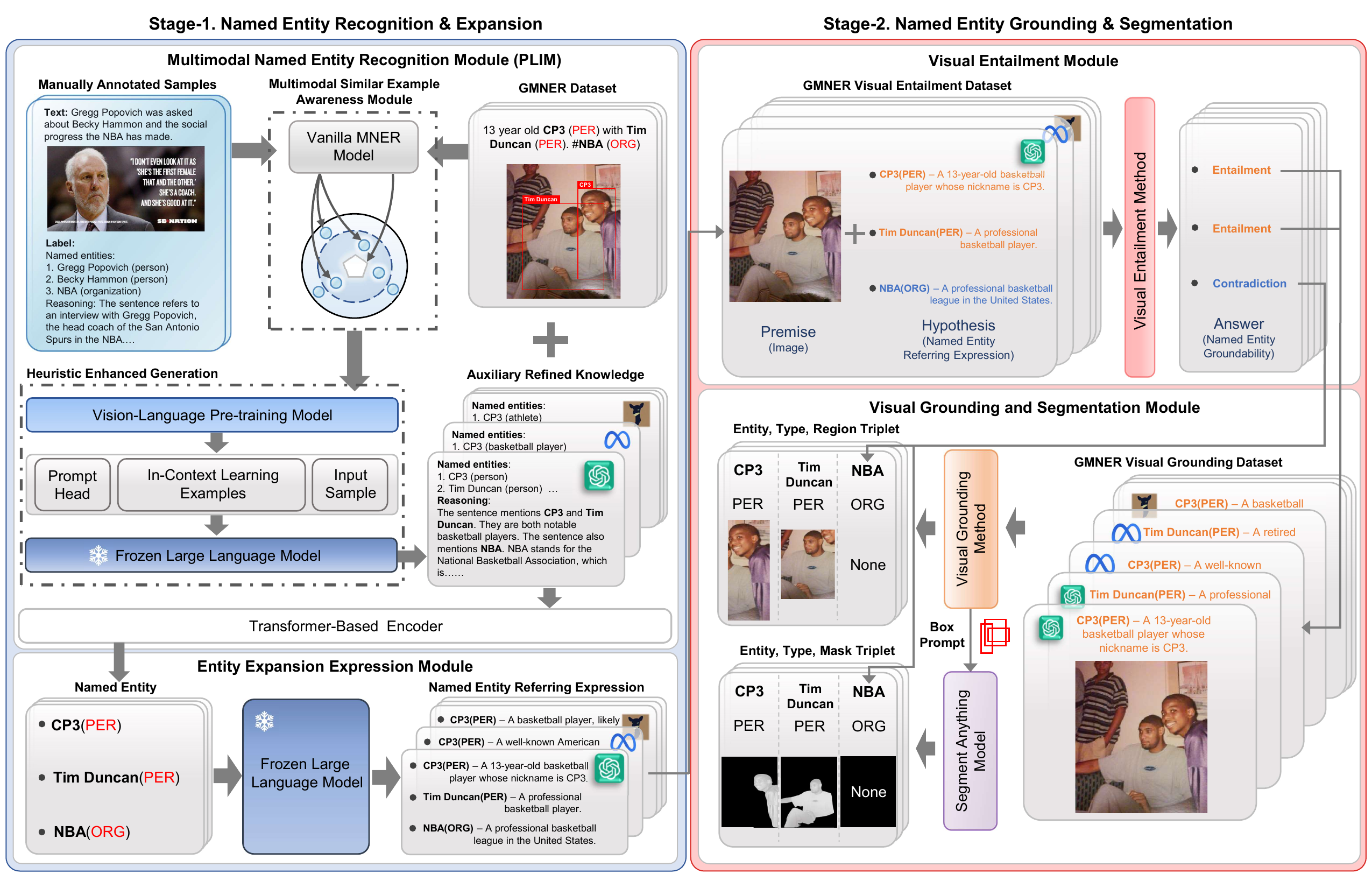}
	\end{center} 
	\caption{The overall architecture of the proposed framework. In the first stage, we create a certain number of manually annotated samples, then heuristically guide LLMs through the Multimodal Similar Example Awareness module to perform in-context learning and generate auxiliary refined knowledge. Additionally, LLMs are utilized to generate named entity referring expressions for text named entities, thereby bridging the gap between text named entities and noun phrases. In the second stage, the Visual Entailment module is responsible for receiving images and named entity referring expressions and determining the visual groundability of named entities. The Visual Grounding and Segmentation module is responsible for performing the final grounding of groundable named entities and providing box prompts to SAM to obtain segmentation masks for visual objects.} 
	\label{fig:framework} 
\end{figure*}

RiVEG mainly consists of two stages, aligned with the pipeline design established in our previous conference paper~\cite{li2023prompting,li2024llms}. The first stage aims to ensure optimal MNER performance and further perform entity expansion expression (detailed in §~\ref{sec:Stage-1}). The second stage aims to reformulate the entire EG task as a joint of VE and VG (detailed in §~\ref{sec:Stage-2}). 
Furthermore, this journal version extends RiVEG by incorporating a new segmentation head in the second stage to support the SMNER task, where bounding-box prompts are used to guide the generation of segmentation masks. In addition, we employ various LLMs to perform direct and effective data augmentation on the overall framework in this extended version. The overall architecture of RiVEG is shown in Fig.~\ref{fig:framework}.

\subsection{Task Formulation}
Given a sentence $\boldsymbol{s} = \{s_1, \cdots, s_{k_{1}}\}$ with $k_{1}$ tokens and its corresponding image $\boldsymbol{v}$. The goal of the MNER task is to recognize and classify all named entities in the sentence $\boldsymbol{s}$. The GMNER task aims to further determine the visually grounded region of each named entity in the image on the basis of the MNER task. And the SMNER task aims to determine the visual segmentation mask of each named entity in the image on the basis of the MNER task. The output of this series of tasks is defined as follows: \vspace{-3pt}
\begin{align}
    Y_{\text{MNER}} &= \{(e_1,t_1), \cdots, (e_{n_1},t_{n_1})\}  \nonumber \\
    Y_{\text{GMNER}} &= \{(e_1,t_1,r_1), \cdots, (e_{n_2},t_{n_2},r_{n_2})\} \nonumber \\
    Y_{\text{SMNER}} &= \{(e_1,t_1,m_1), \cdots, (e_{n_3},t_{n_3},m_{n_3})\}  \nonumber \vspace{-5pt}
\end{align}  
where $e_i$ represents the text named entity, $t_i$ represents the entity type of $e_i$ (\emph{i.e.}, $\textit{PER}$, $\textit{LOC}$, $\textit{ORG}$ and $\textit{MISC}$), $r_i$ represents the visually grounded region of $e_i$, and $m_i$ represents the visual segmentation mask of $e_i$. Note that if $e_i$ has no corresponding visual object in the image, $r_i$ or $m_i$ is specified as $\textit{None}$. Otherwise, $r_i$ comprises 4D coordinates representing the top-left and bottom-right positions of the bounding box, while $m_i$ consists of the binary segmentation mask. For predicted visual region $r_i$ and segmentation mask $m_i$, it is considered correct only if its IoU with a certain gold label is greater than 0.5. 
For the SMNER task, the pixel-level IoU metric is adopted, which directly reflects the overlap between the predicted mask and the ground truth mask. For the GMNER task, the box-level IoU metric is employed, which measures the overlap between the predicted bounding box and the ground truth box. For all three tasks, only predictions where all elements are correct are considered correct. 

\begin{figure}[t] 
	\begin{center}
		\includegraphics[width=1\linewidth]{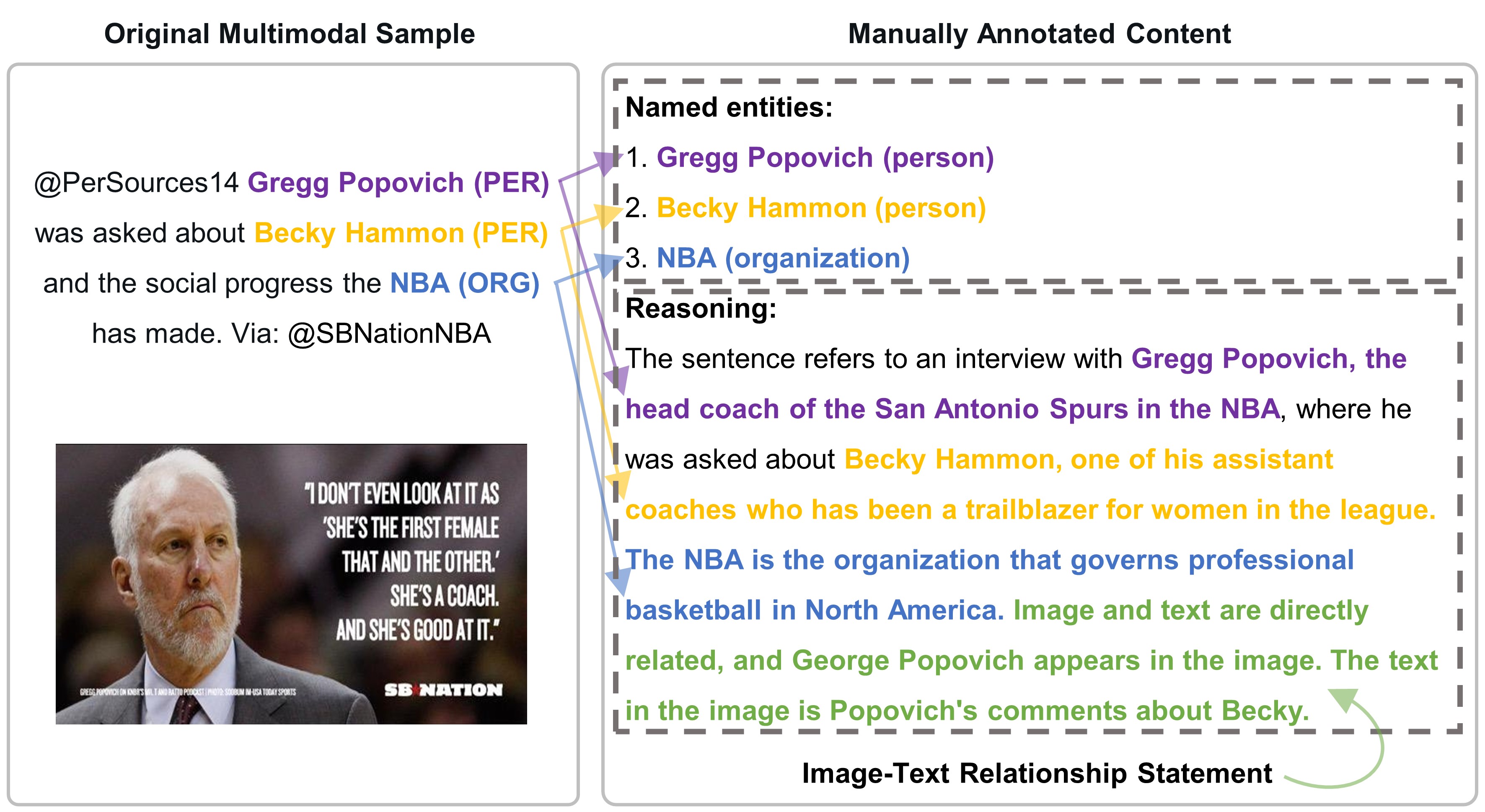}
	\end{center}
	\caption{\textcolor{black}{For a certain number of randomly selected samples, annotation content involves predicting named entities, providing explanations for each entity, and concluding with a statement on the image-text relationship.}}  
	\label{fig:annotated_sample}
\end{figure} 

\subsection{Stage-1. Named Entity Recognition and Expansion} \label{sec:Stage-1}

\subsubsection{Multimodal Named Entity Recognition Module}
Introducing document-level external knowledge into the MNER task is proven to be an effective approach~\cite{wang2022named}. Given that external knowledge retrieved by the existing T-T paradigm MNER method~\cite{wang2022named} is redundant, and LLMs are difficult to be improved with limited MNER data, we regard LLMs as suitable external implicit knowledge bases for the MNER task.

\paragraph{Manually Annotated Samples}
Providing appropriate in-context examples is key to enabling LLMs to perform effective in-context learning~\cite{yang2022empirical,shao2023prompting}. However, this kind of examples that can reflect the way of providing knowledge expansion cannot be simply obtained from the original MNER dataset. To address this challenge, we employ three annotators and randomly select a certain number of samples from the training set of the MNER dataset for manual annotation. Annotators are tasked with evaluating and interpreting the samples from a human perspective. Specifically, the number of annotated samples is 200 for the Twitter-2017 dataset and 120 for the Twitter-2015 dataset. As shown in Fig.~\ref{fig:annotated_sample}, the annotation content mainly consists of two parts. The first part aims to identify the named entities within the sentence, while the second part aims to comprehensively analyze both the image and text content, providing supporting arguments for the identified named entities through the utilization of external knowledge. Additionally, to address the weak correlation between image-text pairs, annotators also need to specify whether named entities are reflected in the image during the annotation process. Through the aforementioned manually annotated samples, we address the challenge of acquiring in-context examples. Such examples can better guide LLMs to imitate and generate more valuable outputs.

\paragraph{Multimodal Similar Example Awareness Module}
We design a Multimodal Similar Example Awareness (MSEA) module to adaptively select relevant manually annotated samples for different input samples. Since the prediction of MNER relies on the joint interaction of textual and visual information, we use the similarity between the multimodal fusion features of samples as the evaluation criterion for similar examples. And these fusion features can be naturally obtained from previous vanilla MNER models. Denote the original MNER dataset as $\mathcal{D_{\text{MNER}}}$, the subset of samples randomly selected from $\mathcal{D_{\text{MNER}}}$ for annotation as $\mathcal{A_{\text{MNER}}}$, and the manually annotated samples dataset as $\mathcal{A_{\text{MNER}}^{'}}$:
\begin{align}
    \mathcal{D_{\text{MNER}}} &= \{(\boldsymbol{s}_1, \boldsymbol{v}_1), \cdots, (\boldsymbol{s}_{m_{1}}, \boldsymbol{v}_{m_{1}})\} \nonumber \\
    \mathcal{A_{\text{MNER}}} &= \{(\boldsymbol{s}_1, \boldsymbol{v}_1), \cdots, (\boldsymbol{s}_{m_{2}}, \boldsymbol{v}_{m_{2}})\} \nonumber \\
    \mathcal{A_{\text{MNER}}^{'}} &= \{(\boldsymbol{s}_1, \boldsymbol{v}_1, \boldsymbol{a}_1), \cdots, (\boldsymbol{s}_{m_{2}}, \boldsymbol{v}_{m_{2}}, \boldsymbol{a}_{m_{2}})\} \nonumber 
\end{align}
where $\boldsymbol{s}_i$ and $\boldsymbol{v}_i$ represent the sentence and image, $\boldsymbol{a}_i$ represents the corresponding manually annotated content. The multimodal encoder $\mathcal{M}_{b}$ of the vanilla MNER model based on dataset $\mathcal{D_{\text{MNER}}}$ is utilized to construct multimodal fusion features for image-text pairs: 
\begin{align}
	\mathcal{F}_{D} = \mathcal{M}_{b}(\mathcal{D_{\text{MNER}}}) \nonumber,
	\mathcal{F}_{A} = \mathcal{M}_{b}(\mathcal{A_{\text{MNER}}}) \nonumber 
\end{align}
where $\mathcal{F}_{D}=\{\boldsymbol{f}_1, \cdots,\boldsymbol{f}_{m_{1}}\}$ and $\mathcal{F}_{A}=\{\boldsymbol{f}_1, \cdots,\boldsymbol{f}_{m_{2}}\}$ represent multimodal fusion features of input samples and annotated samples. Considering that samples with similar features in high-dimensional space have greater pattern similarity, we compute the cosine similarity between the fusion features of any input sample $\boldsymbol{f}_{j} \in \mathcal{F}_{D}$ and all annotated samples $\boldsymbol{f}_{i} \in \mathcal{F}_{A}$: 
\begin{align}
    \mathcal{I}_{j} = \mathop{\text{argTopN}}
    \limits_{i\in\{1,2,...,m_{2}\}} 
    \frac{\boldsymbol{f}_{j}^T \boldsymbol{f}_{i}}{\|\boldsymbol{f}_{j}\|_2\|\boldsymbol{f}_i\|_2}
    \nonumber
\end{align}
where $j\in\{1, \cdots,m_{1}\}$ and $\mathcal{I}_{j}$ is the index set of top-$N$ similar annotated samples of $\boldsymbol{f}_{j}$. For any input sample $(\boldsymbol{s}_j, \boldsymbol{v}_j)\in\mathcal{D_{\text{MNER}}}$, the set of in-context examples $\mathcal{C}_{j}$ used to guide LLMs to generate auxiliary knowledge is defined as: 
\begin{align}
    \mathcal{C}_{j} = \{(\boldsymbol{s}_{i}, \boldsymbol{v}_{i}, \boldsymbol{a}_{i})\ |\ i\in\mathcal{I}_{j}\} \nonumber 
\end{align}
where $(\boldsymbol{s}_i, \boldsymbol{v}_i, \boldsymbol{a}_i)\in\mathcal{A_{\text{MNER}}^{'}}$. Note that the multimodal fusion features of all manually annotated samples can be calculated and stored in advance to achieve more effective awareness of similar examples.

\paragraph{Heuristic Enhanced Generation}
Once the set of in-context examples $\mathcal{C}$ of the input sample is determined with the assistance of the MSEA module, the complete heuristic prompt content is established to activate the ability of LLMs for in-context learning in MNER. For any input sample $(\boldsymbol{s}_j, \boldsymbol{v}_j)\in\mathcal{D_{\text{MNER}}}$, a fixed prompt head, a set of in-context examples $\mathcal{C}_j$, along with the input sample itself, constitute the complete input for LLMs. The prompt head is designed to describe the MNER task using natural language. In-context examples and the input sample are formulated as the following template:

\begin{table}[htbp] \vspace{-1.5mm}
\centering
\resizebox{0.48\textwidth}{!}{%
\begin{tabular}{|l|}
\hline
\textbf{Text}: \textcolor{blue}{$\{Sentence\}$} \\
\textbf{Image}: \textcolor{blue}{$\{Image\ \allowbreak Description\}$} \\
\textbf{Question}: \texttt{Comprehensively analyze the Text and the Image,} \\
\texttt{which named entities and their corresponding types are} \\
\texttt{included in the Text? Explain the reason for your judgment.} \\
\textbf{Answer}: \textcolor{blue}{$\{External\ \allowbreak Annotation\}$} \\
\hline
\end{tabular}}
\end{table} \vspace{-2mm}

\noindent
where \textcolor{blue}{$\{Image\ \allowbreak Description\}$} is the image description of $\boldsymbol{v}_j$ obtained using the multimodal pre-training model, \textcolor{blue}{$\{Sentence\}$} is the original text $\boldsymbol{s}_j$. For the in-context example, \textcolor{blue}{$\{External\ \allowbreak Annotation\}$} is the manually annotated content, while for the new input sample, \textcolor{blue}{$\{External\ \allowbreak Annotation\}$} is empty for LLMs to generate responses. Note that for the same input sample, the same algorithm is applied to different LLMs. This can generate different styles of auxiliary refined knowledge for the same sample and achieve data augmentation.

\paragraph{Entity Prediction Based on Auxiliary Refined Knowledge}
The auxiliary refined knowledge generated by LLMs with $k_{2}$ tokens is defined as $\boldsymbol{z} = \{z_1, \cdots, z_{k_{2}}\}$. The input to the transformer-based encoder is the concatenation of the original text $\boldsymbol{s}$ and $\boldsymbol{z}$: 
\begin{align}
 \text{embed} ([\boldsymbol{s};\boldsymbol{z}]) =
\{h_1,\cdots,h_{k_{1}},\cdots,h_{{k_{1}}+{k_{2}}}\} \nonumber
\end{align}
where $\boldsymbol{h} = \{h_1, \cdots, h_{k_{1}}\}$ is the obtained original text feature integrated with external knowledge. The linear-chain Conditional Random Field (CRF)~\cite{lafferty2001conditional} receives the text feature $\boldsymbol{h}$ and generates the prediction sequence $\boldsymbol{y} = \{y_1, \cdots, y_{k_{1}}\}$: 
\begin{align}
    P(\boldsymbol{y}|\boldsymbol{s},\boldsymbol{z}) = \frac{\prod\limits_{i=1}^{k_{1}} \psi(y_{i-1}, y_i, h_i)}{\sum\limits_{\vy' \in Y} \prod\limits_{i=1}^{k_{1}} \psi(y'_{i-1}, y'_i, h_i)}\nonumber
\end{align}
where $\psi$ is the potential function, $Y$ is the set of all possible label sequences given the input $\boldsymbol{s}$ and $\boldsymbol{z}$. Finally, negative log-likelihood (NLL) as the loss function for the input sequence with gold labels $\vy^*$: 
\begin{align}
\mcL_{\text{MNER-NLL}}(\theta) = - \log P_{\theta}(\vy^*|\boldsymbol{s},\boldsymbol{z}) \nonumber
\end{align}
where $\theta$ is the model parameters. Note that since this step only involves a single text modality, the method here is not unique and can be naturally adapted to any improved knowledge-based NER methods.

\subsubsection{Entity Expansion Expression Module}
To bridge the gap between real-world named entities and noun phrases, we design an Entity Expansion Expression module to guide LLMs in reformulating fine-grained named entities into semantically meaningful coarse-grained expressions. Specifically, we structure the input to LLMs using the following prompt template: 

\begin{table}[htbp] 
\centering
\resizebox{0.46\textwidth}{!}{%
\begin{tabular}{|l|}
\hline
\textbf{Background}: \textcolor{blue}{$\{Image\ \allowbreak Description\}$} \\
    \textbf{Text}: \textcolor{blue}{$\{Sentence\}$} \\
    \textbf{Question}: \texttt{In the context of the provided information, tell me}\\
    \texttt{briefly what is the \textcolor{blue}{$\{Named\ \allowbreak  Entity\}$} in the Text?}\\
    \textbf{Answer}: \textcolor{blue}{$\{Entity\ \allowbreak Expansion\ \allowbreak  Expression\}$} \\
\hline
\end{tabular}} 
\end{table}

\noindent
Here, the \textcolor{blue}{$\{Entity\ \allowbreak Expansion\ \allowbreak Expression\}$} is generated by LLMs. Unlike the generation of auxiliary refined knowledge, which requires dynamic awareness of similar examples due to its length and complexity, the generation of expansion expressions relies on a fixed set of manually constructed in-context examples. This is because the target expansion expressions are short and concise, making dynamic selection less impactful and more computationally expensive.

Finally, we concatenate the named entity, its entity type, and the generated entity expansion expression to form the final named entity referring expression in the format: 
$Named\ \allowbreak Entity\ \allowbreak (Entity\ \allowbreak Type)-Entity\ \allowbreak Expansion\ \allowbreak Expression$. To promote diversity and enable data augmentation, we utilize different LLMs to produce multiple expansion expressions for the same named entity.

\subsection{Stage-2. Named Entity Grounding and Segmentation} \label{sec:Stage-2}

Through the above processing, the entire GMNER task can be naturally redefined as a joint modeling paradigm spanning the perspectives of MNER, VE, and VG. Due to the abundance of existing research on VE~\cite{xie2019visual,li2021align,wang2022ofa} and VG~\cite{zhu2022seqtr,liu2023polyformer,yan2023universal,wang2022ofa}, the method at this stage is also not unique. Note that this study emphasizes the coordination between various modules rather than alterations in the model architecture. Because one of our primary motivations is to explore how to effectively inherit the pre-training foundation of related research in the GMNER task with very limited training data, thereby giving the entire framework unlimited model scalability and data scalability. 

\subsubsection{Visual Entailment Module}
Our experimental results show that when SoTA VG methods are directly fine-tuned using the Twitter-GMNER training set~\cite{yu2023grounded}, these methods achieve Prec@0.5 scores of 21.87\%~\cite{wang2022ofa} and 9.23\%~\cite{zhu2022seqtr} on the Twitter-GMNER test set.\footnote{For the fine-tuning process, we specify the text input as the named entity along with the entity type. For the ungroundable text input, we define the 4D gold label as ($0, 0, 0, 0$).} This demonstrates the difficulty of endowing existing VG methods with the capability to handle the ungroundable text input through direct fine-tuning with limited GMNER data. To address the disparity between VG methods and the EG task, we introduce the Visual Entailment (VE) module. Classic VE task~\cite{xie2019visual} defines the VE dataset as: 
\begin{equation}
    \mathcal{D_{\text{VE}}}=\{(\boldsymbol{v}_1, \boldsymbol{s}_1, \boldsymbol{l}_1), \cdots, (\boldsymbol{v}_{t_{1}}, \boldsymbol{s}_{t_{1}}, \boldsymbol{l}_{t_{1}})\} \nonumber
\end{equation}
where $\boldsymbol{v}_i$ is the image premise, $\boldsymbol{s}_i$ is the text hypothesis, and $\boldsymbol{l}_i$ is the class label. Three class labels $e$, $n$, $c$ are assigned based on the relationship conveyed by $(\boldsymbol{v}_i, \boldsymbol{s}_i)$. Specifically, $e$ (entailment) represents $\boldsymbol{v}_i \vDash \boldsymbol{s}_i$, $n$ (neutral) represents $\boldsymbol{v}_i \nvDash \boldsymbol{s}_i \land \boldsymbol{v}_i \nvDash \neg \boldsymbol{s}_i$, $c$ (contradiction) represents $\boldsymbol{v}_i \vDash \neg \boldsymbol{s}_i$. Similar but distinct, here we define the VE dataset in the GMNER task as: 
\begin{equation}
    \mathcal{D_{\text{GMNER(VE)}}}=\{(\boldsymbol{v}_1, \boldsymbol{s}_1, \boldsymbol{l}_1), \cdots, (\boldsymbol{v}_{t_{2}}, \boldsymbol{s}_{t_{2}}, \boldsymbol{l}_{t_{2}})\} \nonumber
\end{equation}
where $\boldsymbol{v}_i$ is the image, $\boldsymbol{s}_i$ is the named entity referring expression, and $\boldsymbol{l}_i$ is the class label. Since the groundability of each text named entity in the GMNER dataset is explicit, $\mathcal{D_{\text{GMNER(VE)}}}$ retains only two labels $e$ and $c$.\footnote{Here, $\mathcal{D_{\text{GMNER(VE)}}}$ and $\mathcal{D_{\text{SMNER(VE)}}}$ are not distinguished and can be considered the same. This is because the Twitter-SMNER dataset is constructed based on the Twitter-GMNER dataset. There is no significant difference in the groundability of named entities in these two datasets.} 
$e$ represents that $\boldsymbol{s}_i$ can be grounded in $\boldsymbol{v}_i$, and $c$ represents that $\boldsymbol{s}_i$ cannot be grounded in $\boldsymbol{v}_i$. During the training phase, the dataset $\mathcal{D_{\text{GMNER(VE)}}}$ defined above is utilized to fine-tune existing VE models. And during the inference phase, only named entity referring expressions that are considered groundable by the fine-tuned VE model are input to the subsequent VG module. Ungroundable named entity referring expressions are filtered and specified with their entity grounding result as $\textit{None}$.

\subsubsection{Visual Grounding and Segmentation Module}
The introduction of the Entity Expansion Expression module alleviates the differences between real-world named entities and noun phrases. And the introduction of VE module effectively filters ungroundable text inputs. Therefore, the EG task is naturally unified with any existing VG method. Define the subset of all groundable named entity samples in the GMNER dataset as: 
\begin{equation}
    \mathcal{D_{\text{GMNER(VG)}}}=\{(\boldsymbol{v}_1, \boldsymbol{s}_1, \boldsymbol{r}_1), \cdots, (\boldsymbol{v}_{t_{3}}, \boldsymbol{s}_{t_{3}}, \boldsymbol{r}_{t_{3}})\} \nonumber
\end{equation}
where $\boldsymbol{v}_i$ is the image, $\boldsymbol{s}_i$ is the corresponding named entity referring expression, and $\boldsymbol{r}_i$ is the visual grounding region of $\boldsymbol{s}_i$. $\mathcal{D_{\text{GMNER(VG)}}}$ is used to fine-tune the vanilla VG method. Note that during the training phase, if $\boldsymbol{s}_i$ corresponds to multiple different ground truth bounding boxes, we designate the bounding box with the largest area as the only gold label. 

Additionally, because endowing the model with the ability to output precise pixel-level object masks is more challenging than endowing the model with the ability to output visual object 4D coordinates, and the Twitter-SMNER training set contains less than 5k groundable named entities, it is difficult to obtain satisfactory performance by migrating the above processing method for the VG models directly to the Reference Expression Segmentation models~\cite{yan2023universal,zhu2022seqtr}.\footnote{Table~\ref{tab:VS_Module_test} shows the detailed experimental test results.} However, a series of previous works~\cite{li2022box,tian2021boxinst} show that it is feasible to use the box position prompt to obtain the segmentation mask of the main visual object in the box, and previous work such as OpenSeeD~\cite{zhang2023simple} also shows that obtaining the region mask based on the box position prompt is more effective than predicting the region mask directly based on the text query. We instead try to combine the localization capability of the VG method and the zero-shot image segmentation ability of the SAM~\cite{kirillov2023segment}. Specifically, for any input sample $(\boldsymbol{s}_i,\boldsymbol{v}_i)$ in the Twitter-SMNER dataset, the VE method trained on the $\mathcal{D_{\text{GMNER(VE)}}}$ is used to determine the groundability of the named entity
referring expression $\boldsymbol{s}_i$. For $\boldsymbol{s}_i$ determined to be groundable, the VG method trained on $\mathcal{D_{\text{GMNER(VG)}}}$ is used to predict the 4D coordinates of the visual object associated with $\boldsymbol{s}_i$, \emph{i.e.}, ($r_i^{x_1}, r_i^{y_1}, r_i^{x_2}, r_i^{y_2}$). SAM receives the 4D coordinates and treats that coordinates as the box prompt, generating the segmentation mask $\boldsymbol{m}_{i}$ of $\boldsymbol{s}_i$: 
\begin{align}
	\boldsymbol{m}_{i} = \mathcal{SAM}((r_i^{x_1}, r_i^{y_1}, r_i^{x_2}, r_i^{y_2})) \nonumber
\end{align}
The capabilities of these powerful expert models are combined to accomplish complex SMNER task. We also use this combination with existing GMNER methods~\cite{yu2023grounded} to construct three SMNER baselines, and this process requires no additional training or fine-tuning.

\section{Experiments}

\subsection{Experimental Setup}

\begin{table*}[!]
\small
\centering
\caption{Statistics of the datasets used in different modules. $^\dag$ represents the version after data augmentation.}
\resizebox{1.0\textwidth}{!}{%
\begin{tabular}{l|cccccccccc}
\toprule 
\multirow{2}{*}{\textbf{Split}} 
 &
\multicolumn{6}{c|}{\textbf{MNER Module Datasets}}  & \multicolumn{2}{c|}{\textbf{VE Module Datasets}} & \multicolumn{2}{c}{\textbf{VG Module Datasets}}  \\
 &\multicolumn{1}{c}{\textbf{\#Twitter-2015}} &\multicolumn{1}{c}{\textbf{\#Twitter-2015}$^\dag$} &\multicolumn{1}{c}{\textbf{\#Twitter-2017}} &\multicolumn{1}{c}{\textbf{\#Twitter-2017}$^\dag$} &\multicolumn{1}{c}{\textbf{\#Twitter-GMNER}} & \multicolumn{1}{c|}{\textbf{\#Twitter-GMNER}$^\dag$} & {\textbf{\#$\mathcal{D_{\text{GMNER(VE)}}}$}} & \multicolumn{1}{c|}{\textbf{\#$\mathcal{D_{\text{GMNER(VE)}}^\dag}$}} & {\textbf{\#$\mathcal{D_{\text{GMNER(VG)}}}$}} & \multicolumn{1}{c}{\textbf{\#$\mathcal{D_{\text{GMNER(VG)}}^\dag}$}} \\ 
  \midrule 
Train &4000&20000&3373&16865 & 7000 & \multicolumn{1}{c|}{35000} & 11777 & \multicolumn{1}{c|}{58885} & 4733 & 23665\\
Dev  &1000&1000&723&723 &  1500 & \multicolumn{1}{c|}{1500}  & 2447  & \multicolumn{1}{c|}{2447} & 991 & 991\\
Test &3257&3257&723&723&  1500 & \multicolumn{1}{c|}{1500}  & 2542 & \multicolumn{1}{c|}{2542} & 1045 & 1045 \\
\midrule
Total &8257&24257&4819&18311& 10000  &  \multicolumn{1}{c|}{38000} & 16766 &  \multicolumn{1}{c|}{63874} & 6769 & 25701  \\
\bottomrule
\end{tabular}%
}
\label{Different Module Datasets} \vspace{-4mm}
\end{table*}

\subsubsection{Datasets}
Detailed statistics of the datasets used by different modules are shown in Table~\ref{Different Module Datasets}. For the MNER task, we conduct experiments on two classic MNER datasets, \emph{i.e.}, Twitter-2015 and Twitter-2017. Twitter-2015 is constructed by Zhang \textit{et al.}~\cite{zhang2018adaptive}, and Twitter-2017 is constructed by Yu \textit{et al.}~\cite{yu2020improving} based on the SNAP dataset~\cite{lu2018visual}. For the GMNER task, the experiments are based on the currently only Twitter-GMNER dataset~\cite{yu2023grounded}. $\mathcal{D_{\text{GMNER(VE)}}}$ and $\mathcal{D_{\text{GMNER(VG)}}}$ used for training and validating the VE and VG modules are subsets of the Twitter-GMNER dataset. For the SMNER task, the statistical results are based on our proposed Twitter-SMNER dataset (detailed in §~\ref{sec:SMNER_dataset}). Additionally, the base training sets of all datasets are constructed solely based on gpt-3.5-turbo. And 4 different versions of LLMs (\emph{i.e.}, vicuna-7b-v1.5~\cite{chiang2023vicuna}, vicuna-13b-v1.5~\cite{chiang2023vicuna}, llama-2-7b-chat-hf~\cite{touvron2023llama}, llama-2-13b-chat-hf~\cite{touvron2023llama}) are used to implement data augmentation of the training data. These models are widely adopted in related studies~\cite{shi2024generative,mi2024vp,liu2024unimel} and are commonly regarded as representative choices. They span different scales (7B and 13B) and can be run on a single 24GB GPU, striking a balance between performance and reproducibility. All dev and test sets do not involve data augmentation and are constructed based on gpt-3.5-turbo.

\subsubsection{Model Configurations}
For the MNER module, we choose the backbone of UMT~\cite{yu2020improving} as the vanilla MNER model to extract multimodal fusion features. In the Heuristic Enhanced Generation stage, the multimodal pre-training model BLIP-2~\cite{li2023blip} is used to extract short image captions for images. The same BERT$_\text{base}$~\cite{kenton2019bert} and XLM-RoBERTa$_\text{large}$~\cite{conneau2020unsupervised} as in previous MNER research~\cite{yu2020improving,wang2022cat,wang2022ita,wang2022named} are selected as text encoders. 
In the Entity Expansion Expression module, the multimodal pre-training model mPLUG-Owl~\cite{ye2023mplug} is used to generate detailed image descriptions. For the VE module, $\mathcal{D_{\text{GMNER(VE)}}}$ and $\mathcal{D_{\text{GMNER(VE)}}^\dag}$ are used for training OFA$_\text{large(VE)}$~\cite{wang2022ofa} and ALBEF-14M~\cite{li2021align}. For the VG module, $\mathcal{D_{\text{GMNER(VG)}}}$ and $\mathcal{D_{\text{GMNER(VG)}}^\dag}$ are used for training OFA$_\text{large(VG)}$~\cite{wang2022ofa} and SeqTR~\cite{zhu2022seqtr}. The version of SAM used to output segmentation masks is ViT-H-SAM~\cite{kirillov2023segment}.

\subsubsection{Implementation Details}
Without being specified, all training is based on a single 4090 GPU. The best models selected on their respective dev sets are used for evaluation on the test sets. Aligned with previous research~\cite{li2023prompting,li2024llms}, precision (Pre.), recall (Rec.), and F1 score are utilized for evaluating the model performance. 

\begin{table}[t!]
\footnotesize
\setlength\tabcolsep{4pt}
\renewcommand{\arraystretch}{1.2}
\centering
\caption{Performance comparison on two classic MNER datasets. Results of all baseline methods are derived from corresponding papers. $^\dag$ denotes that the text encoder used is BERT$_{\text{base}}$. $^\diamond$ denotes that the text encoder used is XLM-RoBERTa$_{\text{large}}$. The marker \textbf{*} refers to significant test p-value $<$ 0.05 when comparing with ITA and MoRe$_{\text{Text/Image}}$.}
\resizebox{0.48\textwidth}{!}{%
\begin{tabular}{l|cccccc}
\toprule 
\multirow{2}{*}{\textbf{MNER}}
& \multicolumn{3}{c|}{\textbf{Twitter-2015}} & \multicolumn{3}{c}{\textbf{Twitter-2017}} \\
& Pre. & Rec. & \multicolumn{1}{c|}{F1}  & Pre. & Rec. & \multicolumn{1}{c}{F1} \\
 \midrule 
BERT$_{\text{base}}$-CRF$_{\text{(Text only)}}^\dag$~\cite{wang2022cat} & 75.56&  73.88 &  \multicolumn{1}{c|}{74.71} & 86.10 &  83.85& \multicolumn{1}{c}{84.96} 
\\
UMT$^\dag$~\cite{yu2020improving} & 71.67  & 75.23&   \multicolumn{1}{c|}{73.41} & 85.28  & 85.34& \multicolumn{1}{c}{85.31}  \\
UMGF$^\dag$~\cite{zhang2021multi} & 74.49 &  75.21& \multicolumn{1}{c|}{74.85} & 86.54 &  84.50 & \multicolumn{1}{c}{85.51} \\
R-GCN$^\dag$~\cite{zhao2022learning} &73.95 & 76.18 & \multicolumn{1}{c|}{75.00} & 86.72  &87.53 & \multicolumn{1}{c}{87.11}  \\
CAT-MNER$^\dag$~\cite{wang2022cat} & \textbf{76.19} & 74.65 & \multicolumn{1}{c|}{75.41} & 87.04  & 84.97 & \multicolumn{1}{c}{85.99}\\
ITA$^\dag$~\cite{wang2022ita} & - & - & \multicolumn{1}{c|}{75.60} & - & - & \multicolumn{1}{c}{85.72}\\
MoRe$_{\text{Wiki-Text}}^\dag$~\cite{li2023prompting} & 73.31 &  74.43& \multicolumn{1}{c|}{73.86} & 85.92& 86.75 & 86.34 
\\
MoRe$_{\text{Wiki-Image}}^\dag$~\cite{li2023prompting} &73.16&  74.64& \multicolumn{1}{c|}{73.89} & 85.49   & 86.38  & 85.94 
\\
\rowcolor[gray]{0.92} \textbf{PLIM$_{\text{LlaMA2-7B}}^\dag$ (Ours)} & 74.87 & 76.53& \multicolumn{1}{c|}{75.69} & 87.27 & 88.82 & 88.04\\
\rowcolor[gray]{0.92} \textbf{PLIM$_{\text{LlaMA2-13B}}^\dag$ (Ours)} & 74.75& 	77.45& \multicolumn{1}{c|}{76.08} & 	86.74& 	87.89& 	87.31 \\
\rowcolor[gray]{0.92} \textbf{PLIM$_{\text{Vicuna-7B}}^\dag$ (Ours)} & 74.70&  	77.72& \multicolumn{1}{c|}{76.18} & 88.26& 87.34&87.80  \\
\rowcolor[gray]{0.92} \textbf{PLIM$_{\text{Vicuna-13B}}^\dag$ (Ours)} & 74.95 &	76.80& \multicolumn{1}{c|}{75.86} &87.11 & 	87.82& 	87.46  \\
\rowcolor[gray]{0.92} \textbf{PLIM$_{\text{ChatGPT}}^\dag$ (Ours)} & 75.84 & 77.76 & \multicolumn{1}{c|}{76.79} &89.09& 90.08& 89.58  \\
\rowcolor[gray]{0.92} \textbf{PLIM$_{\text{Mix}}^\dag$ (Ours)} & 75.92 & \textbf{78.04}& \multicolumn{1}{c|}{\textbf{76.97*}} & 	\textbf{89.24}&	\textbf{90.19}&	\textbf{89.71*}    \\
\midrule 
XLMR$_{\text{large}}$-CRF$_{\text{(Text only)}}^\diamond$~\cite{li2023prompting} & 76.45&  78.22  & \multicolumn{1}{c|}{77.32} & 88.46 &	90.2 3& \multicolumn{1}{c}{89.34} 
\\
ITA$^\diamond$~\cite{wang2022named} & - & - & \multicolumn{1}{c|}{78.03} & - &- & \multicolumn{1}{c}{89.75} 
\\
MoRe$_{\text{Wiki-Text}}^\diamond$~\cite{wang2022named} & -&- & \multicolumn{1}{c|}{77.91} & - & - & \multicolumn{1}{c}{89.50} 
\\
MoRe$_{\text{Wiki-Image}}^\diamond$~\cite{wang2022named} & - & - & \multicolumn{1}{c|}{78.13} & - & - & \multicolumn{1}{c}{89.82} 
\\
\rowcolor[gray]{0.92} \textbf{PLIM$_{\text{LlaMA2-7B}}^\diamond$ (Ours)} & 78.16 & 79.57 & \multicolumn{1}{c|}{78.86} & 90.12 & 91.19 & 90.66  \\
\rowcolor[gray]{0.92} \textbf{PLIM$_{\text{LlaMA2-13B}}^\diamond$ (Ours)} & 78.37 & \textbf{79.86} & \multicolumn{1}{c|}{79.10} & 89.57 & 92.15 & 90.84 \\
\rowcolor[gray]{0.92} \textbf{PLIM$_{\text{Vicuna-7B}}^\diamond$ (Ours)} & 76.97 & 79.42 & \multicolumn{1}{c|}{78.17} & 89.35 & 91.27 & 90.30  \\
\rowcolor[gray]{0.92} \textbf{PLIM$_{\text{Vicuna-13B}}^\diamond$ (Ours)} & 78.46  & 78.92 & \multicolumn{1}{c|}{78.69} & 90.16  & 90.23 & 90.20 \\
\rowcolor[gray]{0.92} \textbf{PLIM$_{\text{ChatGPT}}^\diamond$ (Ours)} & \textbf{79.21} & 79.45 & \multicolumn{1}{c|}{79.33} & 90.86 & 92.01 & 91.43 \\
\rowcolor[gray]{0.92} \textbf{PLIM$_{\text{Mix}}^\diamond$ (Ours)} & 79.10 & 79.78 & \multicolumn{1}{c|}{\textbf{79.44*}} & \textbf{91.12} & \textbf{92.67}& \textbf{91.89*} \\
\bottomrule
\end{tabular}}
\label{tab:MNER_results} \vspace{-5mm}
\end{table} 

\begin{table*}[!]
\small
\renewcommand{\arraystretch}{1.2}
\centering
\caption{Performance comparison on the Twitter-GMNER and Twitter-SMNER dataset.The models marked with $^\dag$ have undergone data augmentation. $\bigtriangleup$ and $\bigtriangleup^\dag$ indicate the improvement compared with the previous SoTA method H-Index.}
\resizebox{1\textwidth}{!}{%
\begin{tabular}{lcccccccccccccccc}
\toprule
\multirow{2}{*}{\textbf{Methods}}
 & \multicolumn{3}{|c|}{\textbf{GMNER}}  & \multicolumn{3}{c|}{\textbf{MNER}} & \multicolumn{3}{c}{\textbf{EEG}} & \multicolumn{3}{|c}{\textbf{EES}} & \multicolumn{3}{|c}{\textbf{SMNER}}   \\
\cline{2-16}
 & \multicolumn{1}{|c}{\textbf{Pre.}} & \textbf{Rec.} & \multicolumn{1}{c|}{\textbf{F1}} & \textbf{Pre.} & \textbf{Rec.} & \multicolumn{1}{c|}{\textbf{F1}} & \textbf{Pre.} & \textbf{Rec.} & \textbf{F1} & \multicolumn{1}{|c}{\textbf{Pre.}} & \textbf{Rec.} & \textbf{F1} & \multicolumn{1}{|c}{\textbf{Pre.}} & \textbf{Rec.} & \multicolumn{1}{c}{\textbf{F1}}\\
 \midrule
  UMT-RCNN-EVG~\cite{yu2020improving}  & \multicolumn{1}{|c}{49.16} & 51.48 & \multicolumn{1}{c|}{50.29} & 77.89 &79.28 &78.58 & \multicolumn{1}{|c}{53.55} & 56.08 & 54.78 &\multicolumn{1}{|c}{-} & -& \multicolumn{1}{c|}{-} & -& -& - \\
  UMT-VinVL-EVG~\cite{yu2020improving} & \multicolumn{1}{|c}{50.15} & 52.52 & \multicolumn{1}{c|}{51.31} & 77.89 &79.28 &78.58 & \multicolumn{1}{|c}{54.35} & 56.91 & 55.60  &\multicolumn{1}{|c}{-} & -& \multicolumn{1}{c|}{-} & -& -& - \\
  UMGF-VinVL-EVG~\cite{zhang2021multi} & \multicolumn{1}{|c}{51.62} & 51.72 & \multicolumn{1}{c|}{51.67} & 79.02& 78.64& 78.83 & \multicolumn{1}{|c}{55.68} & 55.80 & 55.74  &\multicolumn{1}{|c}{-} & -& \multicolumn{1}{c|}{-} & -& -& - \\
  ITA-VinVL-EVG~\cite{wang2022ita} & \multicolumn{1}{|c}{52.37} & 50.77 & \multicolumn{1}{c|}{51.56} & 80.40& 78.37& 79.37 & \multicolumn{1}{|c}{56.57} & 54.84 & 55.69 &\multicolumn{1}{|c}{54.17} & 52.38& \multicolumn{1}{c|}{53.26} & 50.09& 48.43&	49.25\\
  BARTMNER-VinVL-EVG~\cite{yan2021unified} & \multicolumn{1}{|c}{52.47} & 52.43 & \multicolumn{1}{c|}{52.45} & 80.65& 80.14& 80.39 & \multicolumn{1}{|c}{55.68} & 55.63 & 55.66 &\multicolumn{1}{|c}{53.56} & 53.45 & \multicolumn{1}{c|}{53.51} &50.52 &50.22&	50.37 \\
  H-Index~\cite{yu2023grounded}  & \multicolumn{1}{|c}{56.16} & 56.67 & \multicolumn{1}{c|}{56.41} & 79.37& 80.10& 79.73 & \multicolumn{1}{|c}{60.90} & 61.46 & 61.18 &\multicolumn{1}{|c}{58.91} & 59.54&	\multicolumn{1}{c|}{59.22} & 54.24 & 54.82& 54.53  \\
   \midrule 
 \rowcolor[gray]{0.92} \textbf{RiVEG$_{\text{BERT}}$ (ours)} & \multicolumn{1}{|c}{64.03} & 63.57 & \multicolumn{1}{c|}{63.80} & 83.12& 82.66& 82.89 & \multicolumn{1}{|c}{67.16} & 66.68 & 66.92 & \multicolumn{1}{|c}{64.18} & 63.73 & \multicolumn{1}{c|}{63.96} & 61.09 & 60.66&60.88 \\
 \rowcolor[gray]{0.92} \textbf{RiVEG$_{\text{XLMR}}$ (ours)} & \multicolumn{1}{|c}{65.09} & 66.68 & \multicolumn{1}{c|}{65.88} & 84.80 & 84.23 & 84.51 & \multicolumn{1}{|c}{67.97} & 69.63 & 68.79 &\multicolumn{1}{|c}{64.90} & 66.48&	\multicolumn{1}{c|}{65.68} & 62.10 & 63.61& 62.85 \\
 \rowcolor[gray]{0.92} $\bigtriangleup$  & \multicolumn{1}{|c}{\textcolor{red}{$+$8.93}} & \textcolor{red}{$+$10.01} & \multicolumn{1}{c|}{\textcolor{red}{$+$9.47}} & \textcolor{red}{$+$5.43}& \textcolor{red}{$+$4.13}& \textcolor{red}{$+$4.78} & \multicolumn{1}{|c}{\textcolor{red}{$+$7.07}} & \textcolor{red}{$+$8.17} & \textcolor{red}{$+$7.61} & \multicolumn{1}{|c}{\textcolor{red}{$+$5.99}}& \textcolor{red}{$+$6.94} & \multicolumn{1}{c|}{\textcolor{red}{$+$6.46}} & \textcolor{red}{$+$7.86} & \textcolor{red}{$+$8.79} & \textcolor{red}{$+$8.32} \\
   \midrule
  \rowcolor[gray]{0.92} \textbf{RiVEG$_{\text{BERT}}^\dag$ (ours)} & \multicolumn{1}{|c}{65.10} & 66.96 & \multicolumn{1}{c|}{66.02} & 83.02 & 85.40 & 84.19 & \multicolumn{1}{|c}{68.21} & \textbf{70.18} & 69.18 & \multicolumn{1}{|c}{65.04} & \textbf{66.91} & \multicolumn{1}{c|}{65.96} & 62.08 & 63.86 & 62.95 \\
 \rowcolor[gray]{0.92} \textbf{RiVEG$_{\text{XLMR}}^\dag$ (ours)} & \multicolumn{1}{|c}{\textbf{67.02}} & \textbf{67.10} & \multicolumn{1}{c|}{\textbf{67.06}} & \textbf{85.71}& \textbf{86.16}& \textbf{85.94} & \multicolumn{1}{|c}{\textbf{69.97}} & 70.06 & \textbf{70.01} & \multicolumn{1}{|c}{\textbf{66.75}} & 66.83 & \multicolumn{1}{c|}{\textbf{66.79}} & \textbf{63.88} & \textbf{63.96} & \textbf{63.92}  \\
 \rowcolor[gray]{0.92} $\bigtriangleup^\dag$  & \multicolumn{1}{|c}{\textcolor{red}{$+$10.86}} & \textcolor{red}{$+$10.43} & \multicolumn{1}{c|}{\textcolor{red}{$+$10.65}} & \textcolor{red}{$+$6.34}& \textcolor{red}{$+$6.06}& \textcolor{red}{$+$6.21} & \multicolumn{1}{|c}{\textcolor{red}{$+$9.07}} & \textcolor{red}{$+$8.60} & \textcolor{red}{$+$8.83} & \multicolumn{1}{|c}{\textcolor{red}{$+$7.84}} & \textcolor{red}{$+$7.29} & \multicolumn{1}{c|}{\textcolor{red}{$+$7.57}} & \textcolor{red}{$+$9.64} & \textcolor{red}{$+$9.14} & \textcolor{red}{$+$9.39} \\
\bottomrule
\end{tabular}}
\label{tab:GMNER_Main_Result} \vspace{-4mm}
\end{table*}

\paragraph{MNER Module}
The number of in-context examples used to prompt LLMs is set to 5. The maximum length of text input is 256. We employ the AdamW optimizer~\cite{loshchilov2018decoupled} to minimize the loss function, along with a warmup linear scheduler to control the learning rate. For all MNER experiments with BERT, the batch size is 16, and the learning rate is set to 5e-5 on the Twitter-2015, Twitter-2017, and Twitter-GMNER datasets. The training epoch is set to 20. For all MNER experiments with XLM-RoBERTa, the batch size is 4, with a learning rate of 5e-6 on the Twitter-2015 dataset and 7e-6 on the Twitter-2017 and Twitter-GMNER datasets. The training epoch is set to 25.

\paragraph{VE Module}
For OFA$_\text{large(VE)}$, the learning rate is 2e-5 and the batch size is 32. We employ the trie-based search strategy to constrain the generated labels to the candidate set. The training epoch is 6. For ALBEF-14M, the learning rate is 2e-5 and the batch size is 24. We fine-tune for 5 epochs on a single V100 32G GPU. The remaining details follow the fine-tuning settings of OFA~\cite{wang2022ofa} and ALBEF~\cite{li2021align} for VE. 

\paragraph{VG Module}
For OFA$_\text{large(VG)}$, the learning rate is 3e-5, the batch size is 32 and the training epoch is 10. For SeqTR~\cite{zhu2022seqtr}, as the pre-training weights of SeqTR are not available, we use the checkpoint after pre-training and 5 epochs of fine-tuning on RefCOCO~\cite{yu2016modeling} as initial weights.~\footnote{\url{https://github.com/seanzhuh/SeqTR}} The maximum length of input text is 30, the learning rate is 5e-4, the batch size is 64 and the training epoch is 10. The remaining settings are consistent with the VG fine-tuning strategy of OFA~\cite{wang2022ofa} and SeqTR~\cite{zhu2022seqtr}.

\subsection{Main Results}
\subsubsection{Results on MNER}
Table~\ref{tab:MNER_results} shows the comparison of the proposed MNER module and different baseline methods. For fair comparison, the same set of methods uses the same text encoder. All results are averages of three runs using different random seeds. We present experimental results for different LLMs as well as Mix versions. The datasets used for experiments with individual LLM versions are constructed by the respective LLM. The training set for the Mix version includes data from all 5 LLMs, while the dev and test sets are only constructed by ChatGPT. The results show that LLMs with smaller parameters generally provide lower quality knowledge than LLMs with larger parameters. Additional external knowledge in different styles for the same samples contributes to further enhancing the capabilities of the model. Obviously, the performance of our MNER module surpasses the two SoTA methods MoRe~\cite{wang2022named} and ITA~\cite{wang2022ita}. This indicates that the quality of external knowledge retrieved from LLMs is higher than that retrieved from Wikipedia. We also observe that MoRe performs worse than simple baseline methods in some cases. This is because auxiliary knowledge derived from Wikipedia may be redundant or contain irrelevant content. The above experiments demonstrate the effectiveness of the proposed MNER module.

\subsubsection{Results on GMNER}
Table~\ref{tab:GMNER_Main_Result} shows the comparison between RiVEG and existing GMNER methods. Following Yu \textit{et al.}~\cite{yu2023grounded}, we also count the MNER and Entity Extraction \& Grounding (EEG) results. EEG focuses solely on extracting entity-region pairs without considering entity types. RiVEG achieves significant advantages over baseline methods. Without any data augmentation involved, RiVEG outperforms the previous SoTA method H-index by 9.47\%, 4.78\%, and 7.61\% on all three tasks, respectively. After data augmentation, the lead further extends to 10.65\%, 6.21\%, and 8.83\%. This demonstrates the rationality and effectiveness of the proposed framework. We maximize the performance of MNER to mitigate the error propagation in GMNER task, leverage the characteristics of the VG method to naturally bypass region feature pre-extraction, and address the weak correlation between images and text by leveraging a dedicated VE module. Note that the modular design of RiVEG allows it to benefit from any subsequent independent module research and achieve further performance improvements.

\subsubsection{Results on SMNER} 
Table~\ref{tab:GMNER_Main_Result} also shows the SMNER results of the proposed RiVEG framework. As there are no existing SMNER baselines, we extend three of the most advanced GMNER baseline methods in the same manner, leveraging the localization capability of GMNER methods to prompt SAM in form of boxes. We also count the Entity Extraction \& Segmentation (EES) subtask, which only focuses on the correctness of extracted entity-mask pairs. The experimental results demonstrate the feasibility of combining the visual object localization capability of any GMNER method with the zero-shot image segmentation capability of SAM to accomplish the SMNER task. The performance loss from boxes to masks is minimal, with the F1 score of all methods decreasing by only about 2\%-3\%. Additionally, due to the stronger capabilities in MNER and GMNER of RiVEG, our approach demonstrates significant superiority in the SMNER task. Without involving any data augmentation, RiVEG outperforms the best baseline method H-index by 6.46\% and 8.32\% on EES and SMNER task. After data augmentation, the advantage further expands to 7.57\% and 9.39\%, respectively. 

\subsection{Detailed Analysis}

\begin{table}[t!]
\footnotesize
\setlength\tabcolsep{6pt}
\renewcommand{\arraystretch}{1.2}
\centering
\caption{Ablation study of PLIM module.}
\resizebox{0.49\textwidth}{!}{%
\begin{tabular}{l|cccccc}
\toprule \multirow{2}{*}{\textbf{Categories}}
& \multicolumn{3}{c|}{\textbf{Twitter-2015}} & \multicolumn{3}{c}{\textbf{Twitter-2017}}\\
& Pre. & Rec. & \multicolumn{1}{c|}{F1}  & Pre. & Rec. & \multicolumn{1}{c}{F1}    \\
\midrule
w/o Auxiliary Knowledge & 76.45  & 78.22 & \multicolumn{1}{c|}{77.32}  & 88.46  & 90.23 & 89.34  \\
w/o Manually Annotated Samples & 77.56 & 78.82 & \multicolumn{1}{c|}{78.19}  & 89.63 & 90.50 & 90.06 \\
w/o MSEA$_{N=1}$     & 78.15  & 79.01 & \multicolumn{1}{c|}{78.58}  & 90.49  & 90.82 & 90.65  \\
w/o MSEA$_{N=5}$     & 78.11  & \textbf{79.82} & \multicolumn{1}{c|}{78.95}  & 90.62  & 91.49 & 91.05  \\
w/o MSEA$_{N=10}$     & 78.47  & 79.21 & \multicolumn{1}{c|}{78.84}  & 90.54  & 91.77 & 91.15 \\ 
MSEA$_{N=1}$     & 78.40 & 79.21 & \multicolumn{1}{c|}{78.76} & 89.90 & 91.63 & 90.76 \\
\textbf{MSEA$_{N=5}$}     & \textbf{79.21} & 79.45 & \multicolumn{1}{c|}{\textbf{79.33}} & \textbf{90.86} & 92.01 & \textbf{91.43} \\
MSEA$_{N=10}$      & 78.58 & 79.67 & \multicolumn{1}{c|}{79.12} & 90.54 & \textbf{92.08} & 91.30 \\
\bottomrule
\end{tabular}}
\label{tab:Ablation_PLIM} \vspace{-4mm}
\end{table}

\begin{table*}[!]
\small
\renewcommand{\arraystretch}{1.2}
\centering
\caption{Performance comparison of different visual variants of RiVEG. Variants from the same set have the same MNER results, since keeping the MNER module constant can clearly illustrate the impact of different VE and VG methods on performance.}
\resizebox{1\textwidth}{!}{%
\begin{tabular}{lcccccccccccccccc}
\toprule
\multicolumn{1}{c}{\textbf{Visual Variants of RiVEG}} 
 & \multicolumn{3}{|c|}{\textbf{GMNER}}  & \multicolumn{3}{c|}{\textbf{MNER}} & \multicolumn{3}{c|}{\textbf{EEG}} & \multicolumn{3}{c|}{\textbf{EES}} & \multicolumn{3}{c}{\textbf{SMNER}}   \\
\cline{2-16}
\multicolumn{1}{c}{MNER-VE-VG}
 & \multicolumn{1}{|c}{\textbf{Pre.}} & \textbf{Rec.} & \multicolumn{1}{c|}{\textbf{F1}} & \textbf{Pre.} & \textbf{Rec.} & \multicolumn{1}{c|}{\textbf{F1}} & \textbf{Pre.} & \textbf{Rec.} & \multicolumn{1}{c|}{\textbf{F1}} & \textbf{Pre.} & \textbf{Rec.} & \multicolumn{1}{c|}{\textbf{F1}} & \textbf{Pre.} & \textbf{Rec.} & \multicolumn{1}{c}{\textbf{F1}}\\ 
  \midrule 
  Baseline (H-Index~\cite{yu2023grounded}) & \multicolumn{1}{|c}{56.16} & 56.67 & \multicolumn{1}{c|}{56.41} & 79.37 & 80.10& 79.73 & \multicolumn{1}{|c}{60.90} & 61.46 & \multicolumn{1}{c|}{61.18} &58.91 & 59.54&	\multicolumn{1}{c|}{59.22} & 54.24 & 54.82& 54.53
 \\
 \midrule 
 \multicolumn{16}{c}{ \emph{w/o Data Augmentation (ChatGPT)}}\\
 \midrule
 \rowcolor[gray]{0.92} PLIM-OFA$_\text{large(VE)}$-SeqTR & \multicolumn{1}{|c}{57.06} & 58.46 & \multicolumn{1}{c|}{57.75} & 84.80 & 84.23 & 84.51 & \multicolumn{1}{|c}{59.79} & 61.25 & \multicolumn{1}{c|}{60.51} &57.86 & 59.42&	\multicolumn{1}{c|}{58.63}&	55.23&	56.67&	55.94

 \\ 
 \rowcolor[gray]{0.92} PLIM-ALBEF-14M-SeqTR & \multicolumn{1}{|c}{57.45} & 58.85 & \multicolumn{1}{c|}{58.14} & 84.80 & 84.23 & 84.51 & \multicolumn{1}{|c}{60.25} & 61.72 & \multicolumn{1}{c|}{60.98} & 58.19&	59.72&	\multicolumn{1}{c|}{58.95} & 55.46&	56.90&	56.17 \\
 \rowcolor[gray]{0.92} PLIM-ALBEF-14M-OFA$_\text{large(VG)}$ & \multicolumn{1}{|c}{64.56} & 66.13 & \multicolumn{1}{c|}{65.33} & 84.80 & 84.23 & 84.51 & \multicolumn{1}{|c}{67.47} & 69.12 & \multicolumn{1}{c|}{68.29} & 65.28& 66.88& \multicolumn{1}{c|}{66.07}&	62.37&	63.89&	63.12  \\
 \rowcolor[gray]{0.92} PLIM-OFA$_\text{large(VE)}$-OFA$_\text{large(VG)}$ & \multicolumn{1}{|c}{65.09} & 66.68 & \multicolumn{1}{c|}{65.88} & 84.80 & 84.23 & 84.51 & \multicolumn{1}{|c}{67.97} & 69.63 & \multicolumn{1}{c|}{68.79} & 64.90 & 66.48 &\multicolumn{1}{c|}{65.68} & 62.10 & 63.61& 62.85 \\
 \midrule 
 \multicolumn{16}{c}{ \emph{ Data Augmentation (Mix)}}\\
 \midrule
 \rowcolor[gray]{0.92} PLIM-OFA$_\text{large(VE)}$-SeqTR & \multicolumn{1}{|c}{59.03} & 59.10 & \multicolumn{1}{c|}{59.06} & 85.71 & 86.16 & 85.94 & \multicolumn{1}{|c}{61.85} & 61.93 & \multicolumn{1}{c|}{61.89}& 59.65&	59.81&	\multicolumn{1}{c|}{59.73}	&56.93& 57.05&	56.99 
 \\ 
 \rowcolor[gray]{0.92} PLIM-ALBEF-14M-SeqTR  & \multicolumn{1}{|c}{59.14} & 59.22 & \multicolumn{1}{c|}{59.18} & 85.71 & 86.16 & 85.94 & \multicolumn{1}{|c}{62.01} & 62.09 & \multicolumn{1}{c|}{62.05} & 60.06 & 60.18&	\multicolumn{1}{c|}{60.12} & 57.27 & 57.39&	57.33
 \\ 
 \rowcolor[gray]{0.92} PLIM-ALBEF-14M-OFA$_\text{large(VG)}$  & \multicolumn{1}{|c}{66.12} & 66.20 & \multicolumn{1}{c|}{66.16} & 85.71 & 86.16 & 85.94 & \multicolumn{1}{|c}{68.99} & 69.07 & \multicolumn{1}{c|}{69.03} & 66.32&	66.40&	\multicolumn{1}{c|}{66.36}&	63.53&	63.60&	63.57
 \\ 
 \rowcolor[gray]{0.92} PLIM-OFA$_\text{large(VE)}$-OFA$_\text{large(VG)}$  & \multicolumn{1}{|c}{67.02} & 67.10 & \multicolumn{1}{c|}{67.06} & 85.71 & 86.16 & 85.94 & \multicolumn{1}{|c}{69.97} & 70.06 & \multicolumn{1}{c|}{70.01} &66.75&	66.83&	\multicolumn{1}{c|}{66.79}&	63.88&	63.96&	63.92
 \\ 
 
\bottomrule
\end{tabular}}
\label{tab:other visual variants} \vspace{-2mm}
\end{table*}

\subsubsection{Ablation Study of MNER}
In ablation experiments shown in Table~\ref{tab:Ablation_PLIM}, we remove manually annotated samples and the MSEA module used to guide LLMs respectively. The backbone of this experiment is XLM-RoBERTa$_{\text{large}}$-CRF, and the LLM is gpt-3.5-turbo. \textit{w/o Manually Annotated Samples} means letting the LLM directly provide explanations without any in-context example prompts. \textit{w/o MSEA} means using a fixed set of randomly selected manually annotated examples for prompting. First, the results indicate that the performance can be improved regardless of how the external knowledge is constructed. Secondly, compared with knowledge generated without guidance, the quality of content generated with guidance is higher, and further marginal gains can be achieved by using more relevant examples. But more in-context examples are not always better, as too many in-context examples may introduce noise. 

\subsubsection{Exploration of More Visual Variants of RiVEG}
Table~\ref{tab:other visual variants} shows the performance of more RiVEG visual variants. We additionally select the classic multimodal pre-training method ALBEF~\cite{li2021align} for the VE module and the classic VG method SeqTR~\cite{zhu2022seqtr} for the VG module. The experimental results indicate that even without considering any data augmentation, the weakest variant exhibits highly competitive results compared with the baseline method. After data augmentation, all visual variants outperform the baseline method. The combination of stronger methods typically yields better performance. And we also observe a significant performance gap between SeqTR and OFA, which may be attributed to the text representation method adopted by SeqTR. The GRU vocabulary of SeqTR is constructed solely based on the training and validation data in $\mathcal{D_{\text{GMNER(VG)}}}$ or $\mathcal{D_{\text{GMNER(VG)}}^\dag}$, which may result in difficulties in handling out-of-vocabulary words during its inference process. 

\begin{table*}[!]
\small
\renewcommand{\arraystretch}{1.2}
\centering
\caption{Performance of different text variants of RiVEG. Here we fix the VE module and VG module unchanged. $^\dag$ indicates that the module undergoes data augmentation.}
\resizebox{1\textwidth}{!}{%
\begin{tabular}{lcccccccccccccccc}
\toprule
\multicolumn{1}{c}{\textbf{Text Variants of RiVEG}} 
 & \multicolumn{3}{|c|}{\textbf{GMNER}}  & \multicolumn{3}{c|}{\textbf{MNER}} & \multicolumn{3}{c|}{\textbf{EEG}} & \multicolumn{3}{c|}{\textbf{EES}} & \multicolumn{3}{c}{\textbf{SMNER}}   \\
\cline{2-16}
\multicolumn{1}{c}{MNER-VE-VG}
 & \multicolumn{1}{|c}{\textbf{Pre.}} & \textbf{Rec.} & \multicolumn{1}{c|}{\textbf{F1}} & \textbf{Pre.} & \textbf{Rec.} & \multicolumn{1}{c|}{\textbf{F1}} & \textbf{Pre.} & \textbf{Rec.} & \multicolumn{1}{c|}{\textbf{F1}} & \textbf{Pre.} & \textbf{Rec.} & \multicolumn{1}{c|}{\textbf{F1}} & \textbf{Pre.} & \textbf{Rec.} & \multicolumn{1}{c}{\textbf{F1}}\\ 
  \midrule 
  Baseline (H-Index~\cite{yu2023grounded}) & \multicolumn{1}{|c}{56.16} & 56.67 & \multicolumn{1}{c|}{56.41} & 79.37 & 80.10& 79.73 & \multicolumn{1}{|c}{60.90} & 61.46 & \multicolumn{1}{c|}{61.18} &58.91 & 59.54&	\multicolumn{1}{c|}{59.22} & 54.24 & 54.82& 54.53 
 \\  
 \midrule 
 \multicolumn{16}{c}{ \emph{w/o Data Augmentation (ChatGPT)}}\\
 \midrule
  \rowcolor[gray]{0.92} PLIM$_\text{BERT}$-OFA$_\text{large(VE)}$-OFA$_\text{large(VG)}$ & \multicolumn{1}{|c}{64.03} & 63.57 & \multicolumn{1}{c|}{63.80} & 83.12 & 82.66 & 82.89 & \multicolumn{1}{|c}{67.16} & 66.68 & \multicolumn{1}{c|}{66.92} &64.18&	63.73&	\multicolumn{1}{c|}{63.96}&	61.09&	60.66&	60.88
\\
 \rowcolor[gray]{0.92} PLIM$_\text{XLMR}$-OFA$_\text{large(VE)}$-OFA$_\text{large(VG)}$ & \multicolumn{1}{|c}{65.09} & 66.68 & \multicolumn{1}{c|}{65.88} & 84.80 & 84.23 & 84.51 & \multicolumn{1}{|c}{67.97} & 69.63 & \multicolumn{1}{c|}{68.79}& 64.90&	66.48&	\multicolumn{1}{c|}{65.68}&	62.10&	63.61&	62.85
 \\
 
 \midrule 
 \multicolumn{16}{c}{ \emph{ Data Augmentation (Mix)}}\\
 \midrule 
 \rowcolor[gray]{0.92} PLIM$_\text{BERT}$-OFA$_\text{large(VE)}$-OFA$_\text{large(VG)}$ & \multicolumn{1}{|c}{65.10} & 66.96 & \multicolumn{1}{c|}{66.02} & 83.02 & 85.40 & 84.19 & \multicolumn{1}{|c}{68.21} & \textbf{70.18} & \multicolumn{1}{c|}{69.18}& 65.04&	\textbf{66.91}&	\multicolumn{1}{c|}{65.96}&	62.08&	63.86&	62.95
 \\
 \rowcolor[gray]{0.92} \textbf{PLIM$_\text{XLMR}$-OFA$_\text{large(VE)}$-OFA$_\text{large(VG)}$} &  \multicolumn{1}{|c}{\textbf{67.02}} & \textbf{67.10} & \multicolumn{1}{c|}{\textbf{67.06}} & \textbf{85.71} & \textbf{86.16} & \textbf{85.94} & \multicolumn{1}{|c}{\textbf{69.97}} & 70.06 & \multicolumn{1}{c|}{\textbf{70.01}}& \textbf{66.75}&	66.83&	\multicolumn{1}{c|}{\textbf{66.79}}&	\textbf{63.88}&	\textbf{63.96}&	\textbf{63.92}
\\ 
\midrule 
 \multicolumn{16}{c}{ \emph{w/o Auxiliary Refined Knowledge in MNER}}\\
 \midrule 
   \rowcolor[gray]{0.92} BERT-CRF-OFA$_\text{large(VE)}$-OFA$_\text{large(VG)}$  & \multicolumn{1}{|c}{61.16} & 59.95 & \multicolumn{1}{c|}{60.55} & 78.93 & 77.47 & 78.19 & \multicolumn{1}{|c}{65.33} & 64.04 & \multicolumn{1}{c|}{64.68} & 62.56&	61.33&	\multicolumn{1}{c|}{61.94} & 58.59&	57.44&	58.01
 \\
    \rowcolor[gray]{0.92} BERT-CRF-OFA$_{\text{large(VE)}}^\dag$-OFA$_{\text{large(VG)}}^\dag$  & \multicolumn{1}{|c}{62.16} & 60.94 & \multicolumn{1}{c|}{61.54} & 78.93 & 77.47 & 78.19 & \multicolumn{1}{|c}{66.29} & 64.99 & \multicolumn{1}{c|}{65.63} & 63.40&	62.16&	\multicolumn{1}{c|}{62.77}&	59.39&	58.22&	58.80
 \\     
    \rowcolor[gray]{0.92} XLMR-CRF-OFA$_\text{large(VE)}$-OFA$_\text{large(VG)}$  & \multicolumn{1}{|c}{63.89} & 64.24 & \multicolumn{1}{c|}{64.06} & 82.68 & 83.13 & 82.90 & \multicolumn{1}{|c}{67.45} & 67.82 & \multicolumn{1}{c|}{67.63} & 64.52&	64.87&	\multicolumn{1}{c|}{64.69} & 60.99& 61.33&	61.16
 \\
    \rowcolor[gray]{0.92} XLMR-CRF-OFA$_{\text{large(VE)}}^\dag$-OFA$_{\text{large(VG)}}^\dag$  & \multicolumn{1}{|c}{64.67} & 65.03 & \multicolumn{1}{c|}{64.85} & 82.68 & 83.13 & 82.90 & \multicolumn{1}{|c}{68.11} & 68.49 & \multicolumn{1}{c|}{68.30} & 65.14&	65.50&	\multicolumn{1}{c|}{65.32} & 61.70& 62.04&	61.87
 \\
\bottomrule 
\end{tabular}} \vspace{-5mm}
\label{tab:other text encoder} 
\end{table*}

\subsubsection{Exploration of More Text Variants of RiVEG}
Table~\ref{tab:other text encoder} explores additional text variants of RiVEG. We fix the VE and VG modules, then replace XLM-RoBERTa$_{\text{large}}$ with BERT$_{\text{base}}$ in the MNER module, and remove the external knowledge generated by LLMs. The main conclusions are as follows: 1) The weaker text encoder clearly results in weaker MNER performance, leading to more pronounced error propagation and a decrease in GMNER and SMNER performance. 2) The MNER results demonstrate that the introduction of auxiliary refined knowledge significantly enhances performance. This once again validates the effectiveness of the proposed MNER module. 3) Experiments with the BERT in the \textit{w/o Auxiliary Refined Knowledge} group suggest that even when the MNER performance is nearly identical to that of the baseline, RiVEG still significantly outperforms the baseline in GMNER and SMNER. This underscores the effectiveness of reformulating the overall task. 4) The tests in the \textit{w/o Auxiliary Refined Knowledge} group also validate the effectiveness of data augmentation. When facing similar MNER results, the VE and VG methods enhanced with data augmentation achieve better GMNER and SMNER results. 5) All 14 variants in Table~\ref{tab:other visual variants} and Table~\ref{tab:other text encoder} further demonstrate that combining the localization capability of VG methods and the zero-shot image segmentation capability of SAM can effectively accomplish the SMNER task, with only minimal performance loss.

\begin{figure}[t] 
	\begin{center}
    \includegraphics[width=1\linewidth]{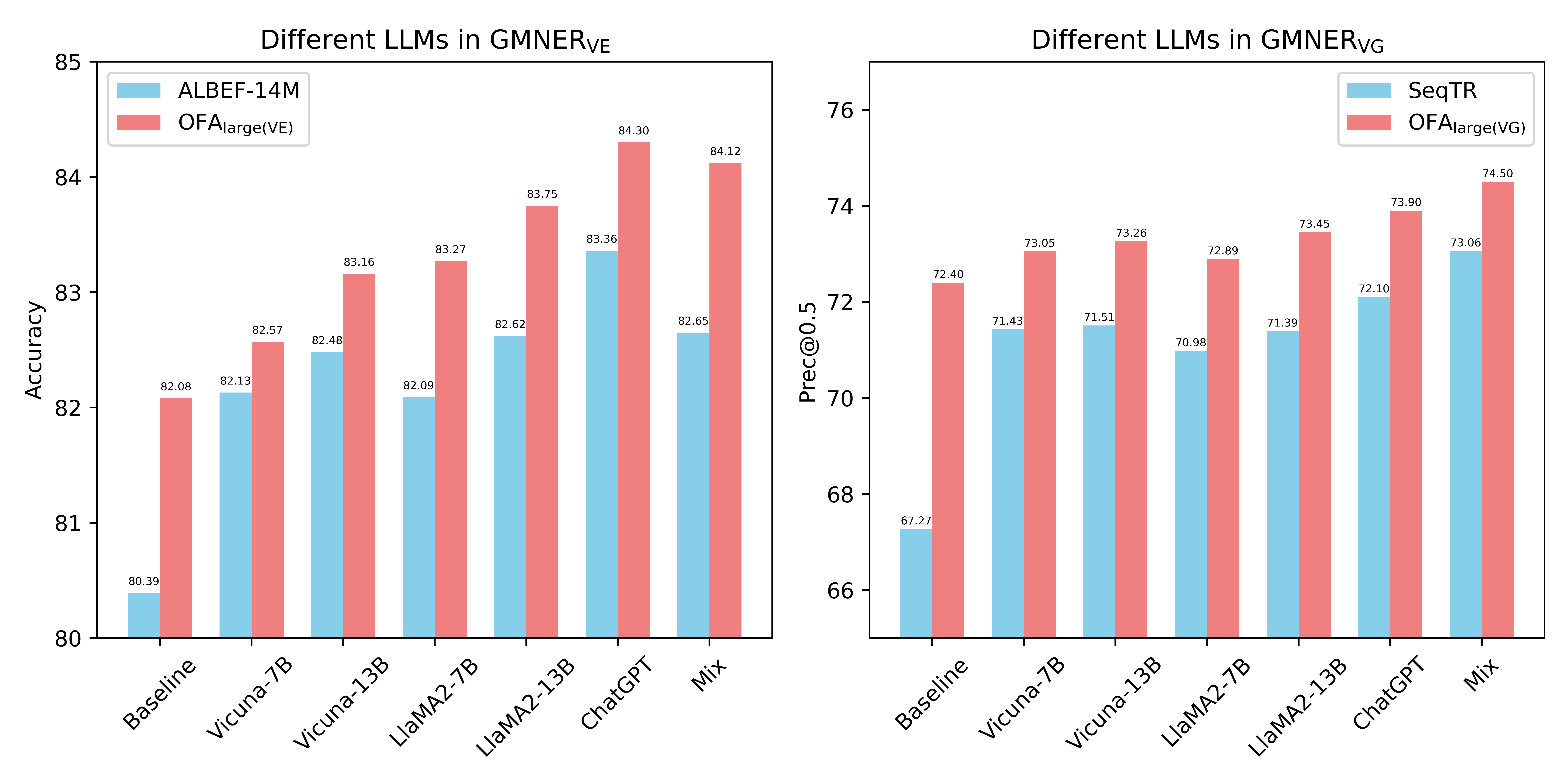}
	\end{center}
	\vspace{-4mm}
	\caption{\textcolor{black}{Exploring entity expansion expressions from various LLMs.}} 
	\label{fig:LLM_in_VE_VG} \vspace{-4mm}
\end{figure}

\subsubsection{Exploration of Entity Expansion Expressions Constructed by Different LLMs}
The named entity expansion expressions generated by different LLMs vary. Fig.~\ref{fig:LLM_in_VE_VG} shows the performance of the same VE and VG models on $\mathcal{D_{\text{GMNER(VE)}}}$ and $\mathcal{D_{\text{GMNER(VG)}}}$ constructed by different LLMs. \textit{Mix} indicates merging the training data of the remaining four LLMs on the basis of the dataset constructed by ChatGPT. The text input of the \textit{Baseline} consists solely of the original named entities and entity types. Clearly, using named entity referring expressions as text inputs are superior to using original named entities. The quality of expansion expressions generated by stronger LLMs is higher than that of weaker LLMs. And mixing more diverse styles of generated data in training data can further enhance performance in some cases.

\subsubsection{Named Entity Referring Expressions Analysis}
Table~\ref{tab:text analysis} shows the performance of VE and VG methods when faced with different types of the text input. The text input of $\mathcal{D_{\text{GMNER(VE)}}}$ and $\mathcal{D_{\text{GMNER(VG)}}}$ constructed from ChatGPT is split into named entities along with entity types, and entity expansion expressions. The results indicate that different methods are suitable for different types of text inputs. Unified frameworks like OFA~\cite{wang2022ofa} are less affected by fine-grained real-world named entities due to their extensive pre-training on a large amount of data. And the impact is more significant for task-specific methods like SeqTR~\cite{zhu2022seqtr}. Different methods exhibit different characteristics, but the proposed hybrid named entity referring expressions, which balance between coarse-grained and fine-grained representations, yield the best performance across all methods.

\begin{table}[t!]
\footnotesize
\setlength\tabcolsep{8pt}
\renewcommand{\arraystretch}{1.1}
\centering
\caption{Ablation study of named entity referring expressions.}
\resizebox{0.4\textwidth}{!}{%
\begin{tabular}{l|cccccc}
\toprule
\multirow{1}{*}{\textbf{VE Module Text Input (Acc.)}} & \multicolumn{1}{c}{\textbf{ALBEF}} & \multicolumn{1}{c}{\textbf{OFA$_\text{large(VE)}$}} \\
\midrule
 Named Entity + Entity Type& 80.39 & 82.08 \\
 Entity Expansion Expression& 81.76 & 81.40 \\
 Named Entity Referring Expression& \textbf{83.36} & \textbf{84.30}  \\\midrule \midrule
\multirow{1}{*}{\textbf{VG Module Text Input (Prec@0.5)}} & \multicolumn{1}{c}{\textbf{SeqTR}} & \multicolumn{1}{c}{\textbf{OFA$_\text{large(VG)}$}} \\
\midrule
 Named Entity + Entity Type& 67.27 & 72.40 \\
 Entity Expansion Expression& 71.72 & 71.10  \\
 Named Entity Referring Expression& \textbf{72.10} & \textbf{73.90}  \\
\bottomrule
\end{tabular}}
\label{tab:text analysis} 
\end{table}

\begin{table}[t!]
\footnotesize
\setlength\tabcolsep{18pt}
\renewcommand{\arraystretch}{1.0}
\centering
\caption{Comparison of two different methods of obtaining segmentation masks.}
\resizebox{0.35\textwidth}{!}{%
\begin{tabular}{l|cccccc}
\toprule
\multirow{1}{*}{\textbf{Segmentation Method}} & \multicolumn{1}{c}{\textbf{mIoU}}  \\
\midrule 
\multicolumn{2}{c}{Text Query-Based}\\
 \midrule
 SeqTR~\cite{zhu2022seqtr} & 56.34 \\
 UNINEXT-R50~\cite{yan2023universal} & 58.66  \\
 \midrule 
\multicolumn{2}{c}{Box Prompt-Based}\\
 \midrule
 OFA$_\text{large(VG)}$~\cite{wang2022ofa} + ViT-H-SAM~\cite{kirillov2023segment} & \textbf{62.12} \\
\bottomrule
\end{tabular}}
\label{tab:VS_Module_test} \vspace{-4mm}
\end{table}

\subsubsection{Comparison Between Text Query-Based and Box Prompt-Based Segmentation Masks Prediction}
Table~\ref{tab:VS_Module_test} shows the comparison between two different methods of obtaining segmentation masks. Specifically, the training data for the two text-query based methods consist of named entity referring expression-mask pairs. These methods directly output segmentation masks based on the input text queries. And the training data for the box prompt-based method consists of named entity referring expression-box pairs. This method first predicts the box corresponding to the text query through visual grounding, and then prompts SAM to obtain the segmentation mask of the main visual object within the box. The results show that the two-stage box prompt-based method outperforms the single-stage text query-based method significantly. This may be attributed to the insufficient training data in the SMNER dataset, which limits the performance of text query-based segmentation methods. VG methods have lower requirements on the amount of data, and the emergence of SAM makes it possible to obtain the segmentation mask of the main visual object from boxes. Therefore, at the current stage, the idea of obtaining segmentation masks based on box prompts may be more suitable for the SMNER task.

\begin{figure}[t] 
	\begin{center}
    \includegraphics[width=1\linewidth]{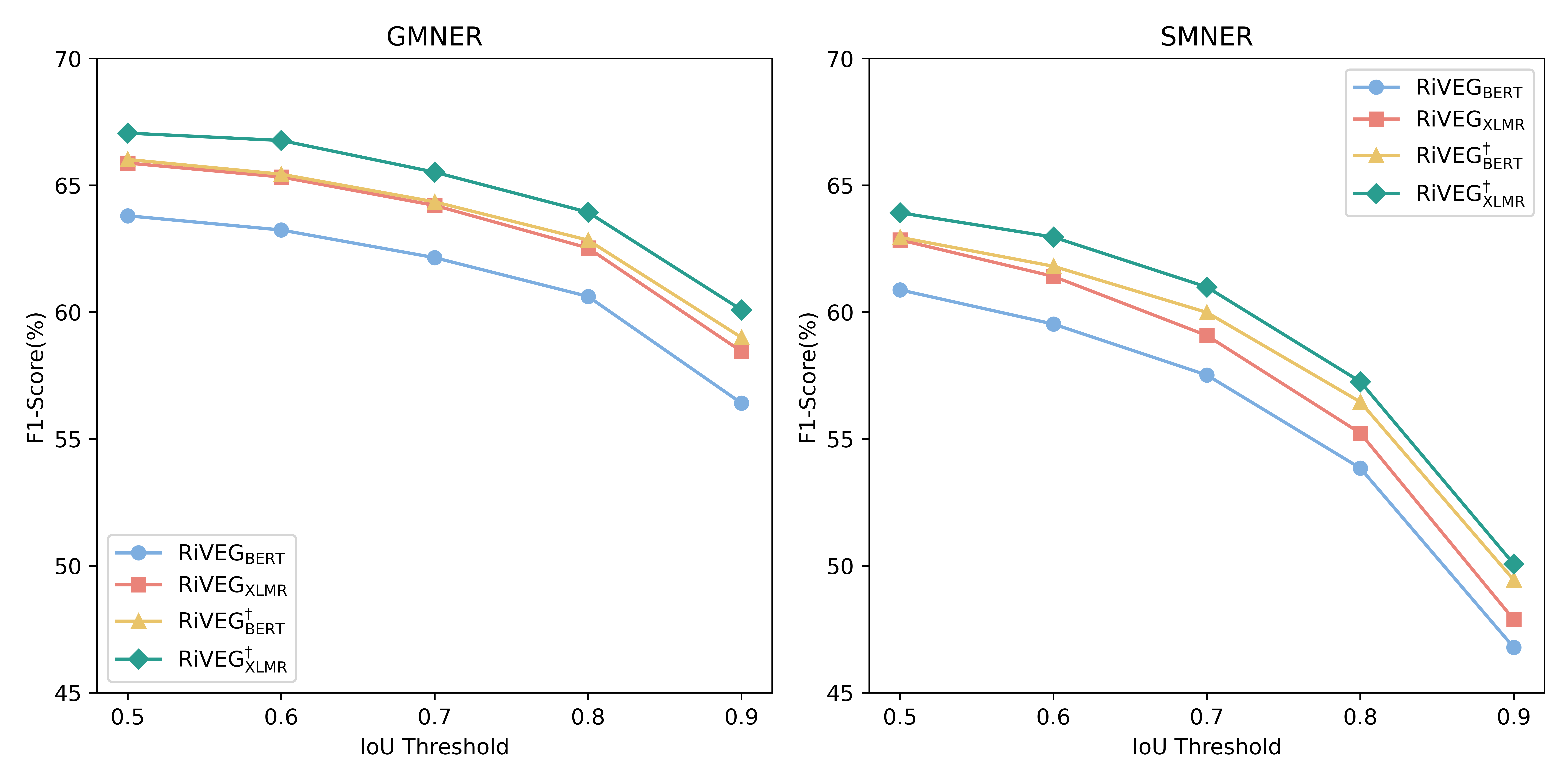}
	\end{center} \vspace{-4mm}
	\caption{\textcolor{black}{The impact of different IoU thresholds on GMNER and SMNER tasks.}} 
	\label{fig:line_chart} 
\end{figure}

\begin{figure}[t] 
	\begin{center}
    \includegraphics[width=1\linewidth]{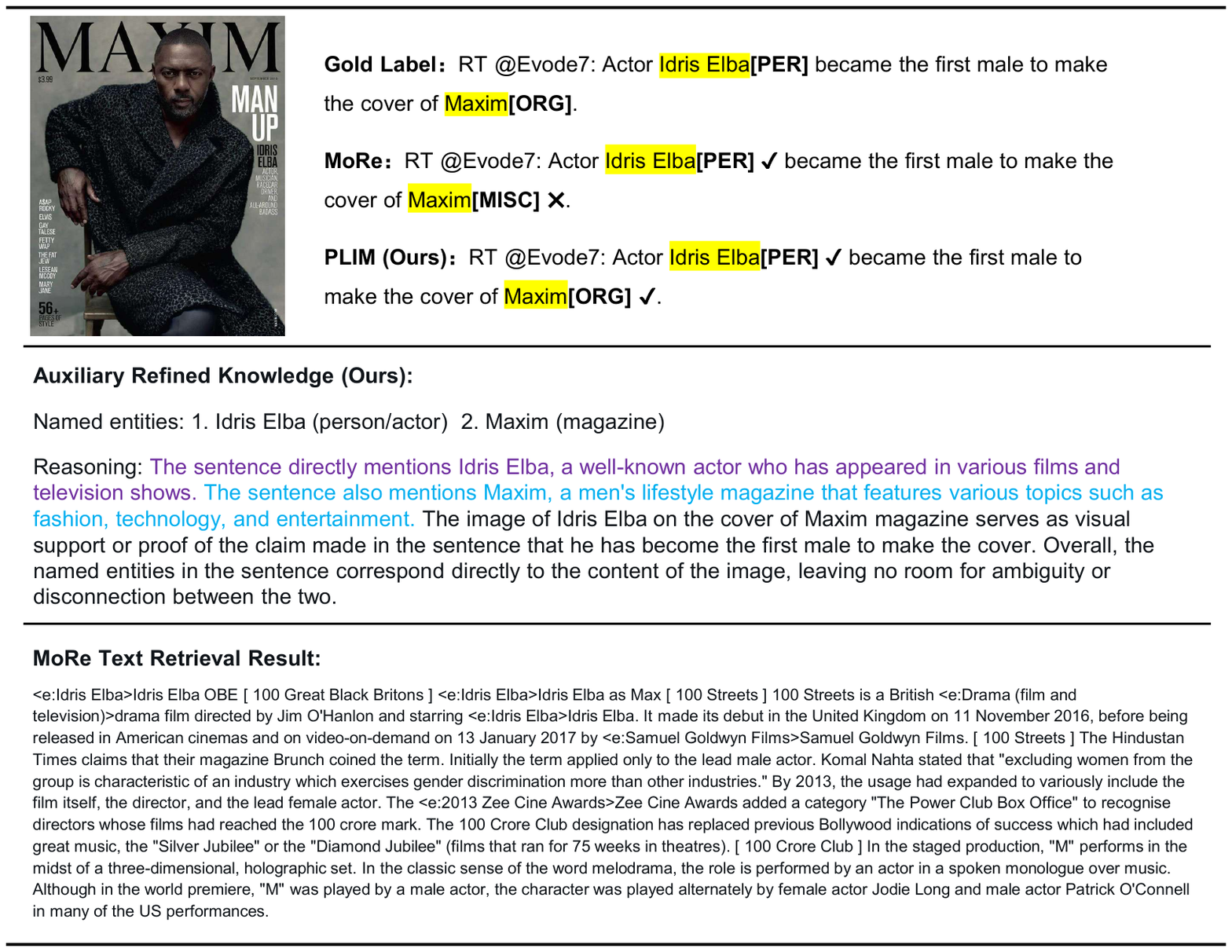}
	\end{center}
	\vspace{-5mm}
	\caption{A case study on how information obtained through different methods affects MNER predictions.} 
	\label{fig:MNER_case_study} \vspace{-6mm}
\end{figure}

\begin{figure}[t] 
	\begin{center}
    \includegraphics[width=1\linewidth]{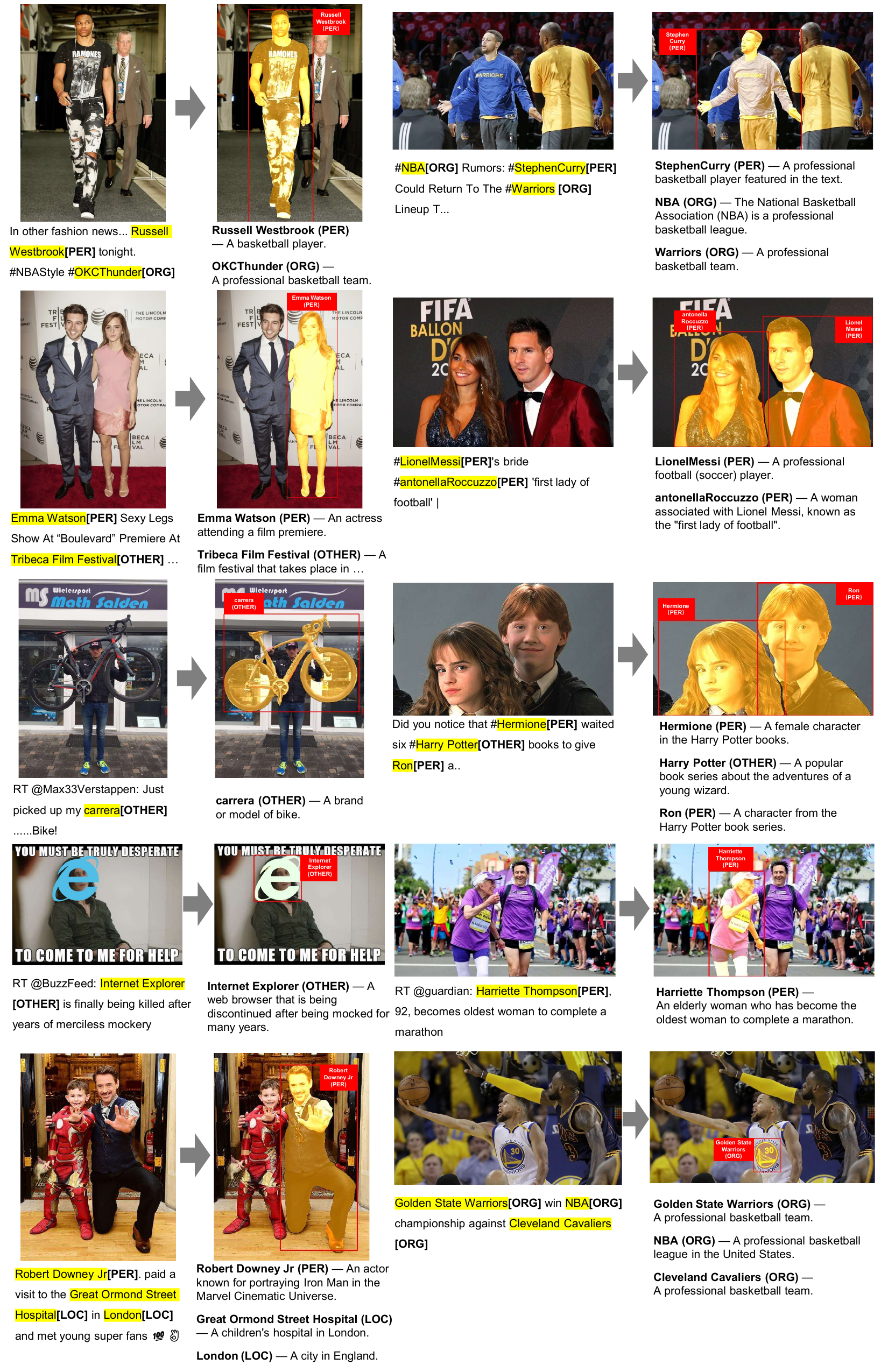}
	\end{center}
	\vspace{-5mm}
	\caption{Qualitative examples of the proposed RiVEG. For each original image-text pair, the subsequent image shows the grounding and segmentation results of the named entities. The text below the image presents the predicted named entities and the constructed expansion expressions.}
	\label{fig:G(S)MNER_case_study}  
\end{figure}

\begin{figure}[t] 
	\begin{center}
    \includegraphics[width=1\linewidth]{
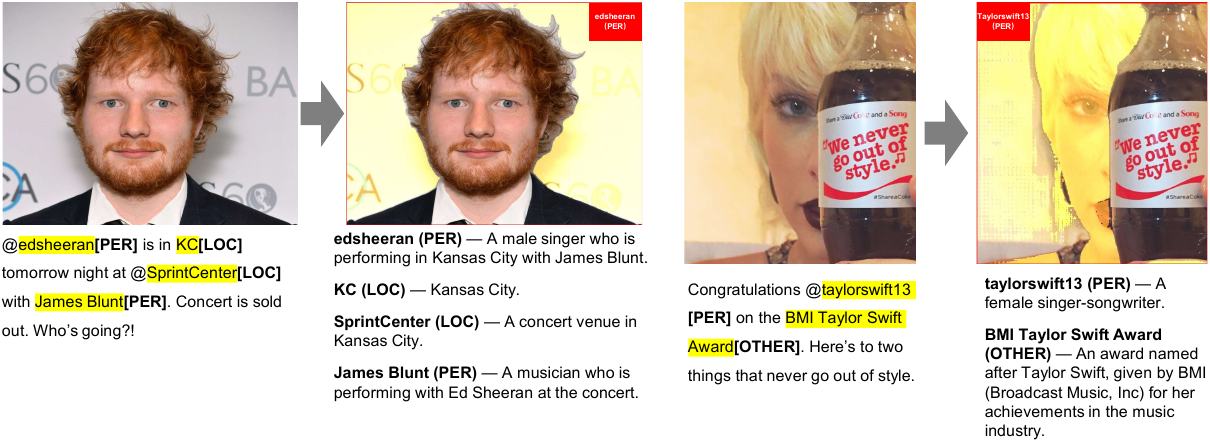}
	\end{center}
	\vspace{-4mm}
	\caption{A case study on two incorrect predictions. The box prompt-based segmentation method also has limitations. In rare cases, the segmentation mask may deviate from the expected visual object within the bounding box.}
	\label{fig:G(S)MNER_bad_case_study}  
  \vspace{-3mm}
\end{figure}

\subsubsection{Sensitivity Analysis of IoU Threshold}
We also evaluate the performance of four RiVEG variants in Table~\ref{tab:GMNER_Main_Result} under different IoU thresholds for the GMNER and SMNER tasks. Fig.~\ref{fig:line_chart} illustrates that as the IoU threshold increases, the performance of all variants gradually decreases. The performance of GMNER is less affected by the increase in IoU threshold compared with SMNER. Additionally, the comparison between RiVEG$_{\text{BERT}}^\dag$ and RiVEG$_{\text{XLMR}}$ further demonstrates the effectiveness of data augmentation. After data augmentation, the performance of RiVEG$_{\text{BERT}}^\dag$ is nearly identical to that of RiVEG$_{\text{XLMR}}$ without data augmentation at the 0.5 threshold, indicating that data augmentation enhances the overall capability of the framework. Furthermore, RiVEG$_{\text{BERT}}^\dag$ with data augmentation is more adapted to higher IoU thresholds, indicating that data augmentation endows the VG module with more accurate visual localization capability.

\subsubsection{Case Study}
Fig.~\ref{fig:MNER_case_study} illustrates how externally acquired knowledge impacts MNER performance. In this case, the knowledge retrieved from Wikipedia may not be closely related to the original sample, introducing unnecessary noise to the model. However, the external knowledge generated by guiding LLMs provides simple and direct explanations tailored to the original sample, effectively assisting the model in making correct predictions. 

In Fig.~\ref{fig:G(S)MNER_case_study}, we provide some case studies to intuitively demonstrate RiVEG prediction performance on challenging samples. These predictions reflect several characteristics: 1) RiVEG can accurately assess the groundability of predicted entities and correctly locate and segment groundable entities. And even in the presence of multiple potential objects in the image, our method can select the correct one, \textit{e.g.,} \textit{Russell Westbrook}, \textit{Robert Downey Jr}, and \textit{Emma Watson}. 2) When multiple groundable named entities appear in the same image, RiVEG predictions are all correct without confusion, \textit{e.g.,} \textit{Hermione} \& \textit{Ron}, \textit{LionelMessi} \& \textit{antonellaRoccuzzo}. This may be attributed to our design of entity expansion expressions accurately describing the characteristics of entities, such as ``\textit{Hermione (PER) -- A female character}" and ``\textit{antonellaRoccuzzo (PER) -- A woman 
associated with Lionel Messi}". These features contribute to more accurate localization of the corresponding visual objects. 3) The performance on entities beyond \textit{PER} type is also excellent, \textit{e.g.,} \textit{carrera}, \textit{Internet Explorer}. 4) The segmentation masks obtained through box prompts are accurate in most cases. 

Despite the overall effectiveness of the box prompt-based segmentation approach, we observe a few rare failure cases where the predicted segmentation mask deviates from the expected visual object within the bounding box. As shown in Fig.~\ref{fig:G(S)MNER_bad_case_study}, the model erroneously captures background regions or unrelated visual content. These examples reveal the limitations of box prompt-based segmentation under complex visual conditions. Although such errors are infrequent and do not significantly impact the overall performance, they highlight promising directions for future work, such as incorporating more adaptive or confidence-aware segmentation strategies.

\section{Research Outlook}

 As illustrated in Fig.~\ref{fig:future outlook}, related research suggests two potentially promising paradigms for the SMNER task: 
 
 1) The end-to-end paradigm leverages multimodal large language models to directly process multimodal inputs and yield structured outputs comprising the entity, its type, and corresponding segmentation. This paradigm offers potential advantages in unified optimization and task generalization, and notably, it is inherently free from error propagation between sequential modules. However, it typically requires substantial training resources and large-scale annotated data. Potential future directions for this paradigm include effective data augmentation strategies, optimized model architectures, and the exploration of using visual entity signals to enhance textual entity recognition in a reverse manner.
 
  2) The cascading paradigm decomposes the task into multiple stages: first, the MNER module is applied to recognize textual entities, followed by diverse visual segmentation modules to localize entity masks. This design enables better modularity and interpretability, although it often introduces architectural complexity and potential error propagation across modules. Future directions include improving module capabilities and simplifying the architecture by reducing component count to mitigate error propagation.

\begin{figure}[t] 
	\begin{center}
    \includegraphics[width=1\linewidth]{future_outlook.jpg}
	\end{center}
	\vspace{-3mm}
	\caption{Two potentially paradigms for the SMNER task.}
	\label{fig:future outlook}  
  \vspace{-3mm}
\end{figure}

This work adopts the cascading design, which demonstrates controllability and explainability. In the future, with the continued advancement of multimodal foundation models, task-specific end-to-end frameworks tailored for fine-grained entity-level understanding may become increasingly feasible, offering a compelling direction for further exploration.

\section{Conclusion}
In this paper, we propose a unified framework RiVEG that spans across MNER, GMNER, and SMNER tasks. 
This framework aims to reformulate the entire task as a joint stage of named entity recognition \& expansion and named entity grounding \& segmentation. This reformulation naturally addresses two major limitations of existing GMNER methods and endows the framework with unlimited data and model scalability by unifying VG and EG. Additionally, we construct a new SMNER task and the corresponding Twitter-SMNER dataset to achieve finer-grained multimodal information extraction. Our practical experiments demonstrate the feasibility of utilizing SAM to enhance any GMNER model for accomplishing the SMNER task. And extensive experiments demonstrate that RiVEG significantly outperforms SoTA methods on the four datasets across three tasks. This work does not focus on modifications to the model architectures within modules, but emphasizes the coordination and complementarity between different modules. We hope that RiVEG can serve as a solid baseline to inspire any related research, ultimately leading to better solutions for these tasks.

\bibliographystyle{IEEEtran}
\bibliography{ref_short}
\end{document}